\documentclass{article}
\usepackage[utf8]{inputenc}
\usepackage{graphicx}
\usepackage{amsmath}
\usepackage{amssymb}
\usepackage{amsthm}
\usepackage[margin=1in]{geometry}
\usepackage{authblk}
\usepackage{color, lineno}

\usepackage{lscape}

\usepackage{hyperref}
\usepackage{cite}

\newcommand{\ww}{\mathrm{WW}}
\newcommand{\wm}{\mathrm{WM}}
\newcommand{\mw}{\mathrm{MW}}
\newcommand{\mm}{\mathrm{MM}}

\begin{document}

\title{Evolutionary stability of cooperation in indirect reciprocity under noisy and private assessment}
\author[a,b,c,1]{Yuma Fujimoto}
\author[a,d]{Hisashi Ohtsuki}

\affil[a]{Research Center for Integrative Evolutionary Science, SOKENDAI (The Graduate University for Advanced Studies). Shonan Village, Hayama, Kanagawa 240-0193, Japan}
\affil[b]{Universal Biology Institute (UBI), the University of Tokyo. 7-3-1 Hongo, Bunkyo-ku, Tokyo 113-0033, Japan.}
\affil[c]{CyberAgent, Inc.. 40-1, Udagawa-cho, Shibuya-ku, Tokyo 150-0042, Japan.}
\affil[d]{Department of Evolutionary Studies of Biosystems, SOKENDAI. Shonan Village, Hayama, Kanagawa 240-0193, Japan.}
\affil[1]{{\rm fujimoto\_yuma@soken.ac.jp}}

\date{}
\maketitle

\begin{abstract}
Indirect reciprocity is a mechanism that explains large-scale cooperation in humans. In indirect reciprocity, individuals use reputations to choose whether or not to cooperate with a partner and update others' reputations. A major question is how the rules to choose their actions and the rules to update reputations evolve. In the public reputation case, where all individuals share the evaluation of others, social norms called Simple Standing (SS) and Stern Judging (SJ) have been known to maintain cooperation. However, in the case of private assessment where individuals independently evaluate others, the mechanism of maintenance of cooperation is still largely unknown. This study theoretically shows for the first time that cooperation by indirect reciprocity can be evolutionarily stable under private assessment. Specifically, we find that SS can be stable, but SJ can never be. This is intuitive because SS can correct interpersonal discrepancies in reputations through its simplicity. On the other hand, SJ is too complicated to avoid an accumulation of errors, which leads to the collapse of cooperation. We conclude that moderate simplicity is a key to success in maintaining cooperation under the private assessment. Our result provides a theoretical basis for evolution of human cooperation.
\end{abstract}

\section{Introduction}
Cooperation benefits others but is costly to the cooperator itself. Nevertheless, cooperation is widespread from microscopic to macroscopic scales, such as among microorganisms, animals, humans, and nations. One way to sustain cooperation is that agents conditionally cooperate with others who cooperate with them, which is realized by, for example, repeated interactions~\cite{trivers1971evolution,axelrod1981evolution, axelrod1984evolution} and partner choice~\cite{yamagishi1984prisoner,noe1994biological,noe1995biological,barclay2013strategies}. Such conditional cooperation based on personal experiences is applicable only to a small population where members can interact directly and repeatedly with most of the others.

However, cooperative behavior is observed even in a large-scale society (e.g., human societies). Since individuals inevitably encounter strangers there, they need reputations of those strangers in order not to cooperate unconditionally. Only individuals with good reputations can receive cooperation. The system that individuals indirectly reward others via their reputations as described above is called indirect reciprocity~\cite{alexander1987biology,nowak1998evolution,nowak2005evolution}. In reality, humans are particularly interested in reputations and gossip about themselves and others~\cite{emler1994gossip, dunbar1998grooming, dunbar2004gossip}. Furthermore, many experiments have pointed out that gossips concern cooperative behaviors 
\cite{feinberg2012virtues, feinberg2014gossip, wu2016reputation}.

Errors that inevitably occur in actions and in assessment hinder cooperation by indirect reciprocity. Indeed, the simplest social norm called image scoring~\cite{nowak1998evolution, nowak2005evolution} fails to maintain full cooperation under errors~\cite{panchanathan2003tale,sigmund2010calculus} (a similar failure is also seen in direct reciprocity~\cite{nowak1993strategy, wu1995cope,sigmund2010calculus}). This is because one erroneous defection triggers further defection. Nevertheless, previous studies have theoretically shown that cooperation can be maintained by the so-called ``leading eight'' social norms~\cite{ohtsuki2004should,ohtsuki2006leading} even in the presence of such errors when all individuals share the reputation of the same individual (i.e., public assessment). Public reputation cases have been thoroughly studied for about two decades~\cite{pacheco2006stern, suzuki2007evolution, santos2007multi, fu2008reputation, suzuki2008evolutionary, uchida2010competition, ohtsuki2015reputation, santos2016social, santos2016evolution, sasaki2017evolution, santos2018social1, santos2018social2, xia2020effect, santos2021complexity, podder2021local}. When individuals cannot share their evaluations of the same target (i.e., private assessment), however, errors cast a shadow over cooperation more crucially. In this case, a single disagreement in opinions between two individuals can lead to further disagreements \cite{uchida2010effect,uchida2013effect,okada2017tolerant,okada2018solution,hilbe2018indirect}. Whether cooperation is maintained under the noisy and private assessment is still largely unsolved in theory and is one of the major open problems in studies of indirect reciprocity \cite{bowles2011cooperative, okada2020review, santos2021complexity}.

Previous studies have shown that maintaining cooperation with indirect reciprocity is very difficult under noisy and private assessment. For example, Hilbe et al.~\cite{hilbe2018indirect} showed by an evolutionary simulation that the above leading eight strategies cannot succeed in cooperation under private assessment. Some studies~\cite{yamamoto2017norm, lee2021local, lee2022second} have demonstrated the emergence of cooperation under noisy and private assessment, but under the restrictive assumption that only local mutations in the strategy space are allowed, thus excluding the possibility that a fully cooperative strategy is directly invaded by free-riders. Other studies have shown that a mechanism to synchronize opinions between individuals has a positive influence on cooperation in indirect reciprocity, such as empathy, generosity, spatial structure, and so on~\cite{brush2018indirect, whitaker2018indirect, radzvilavicius2019evolution, krellner2020putting, quan2020withhold, krellner2021pleasing, schmid2021evolution, kessinger2022indirect, quan2022keeping, gu2022reputation}.

\begin{figure}%[tbhp]
\centering
\includegraphics[width=0.6\linewidth]{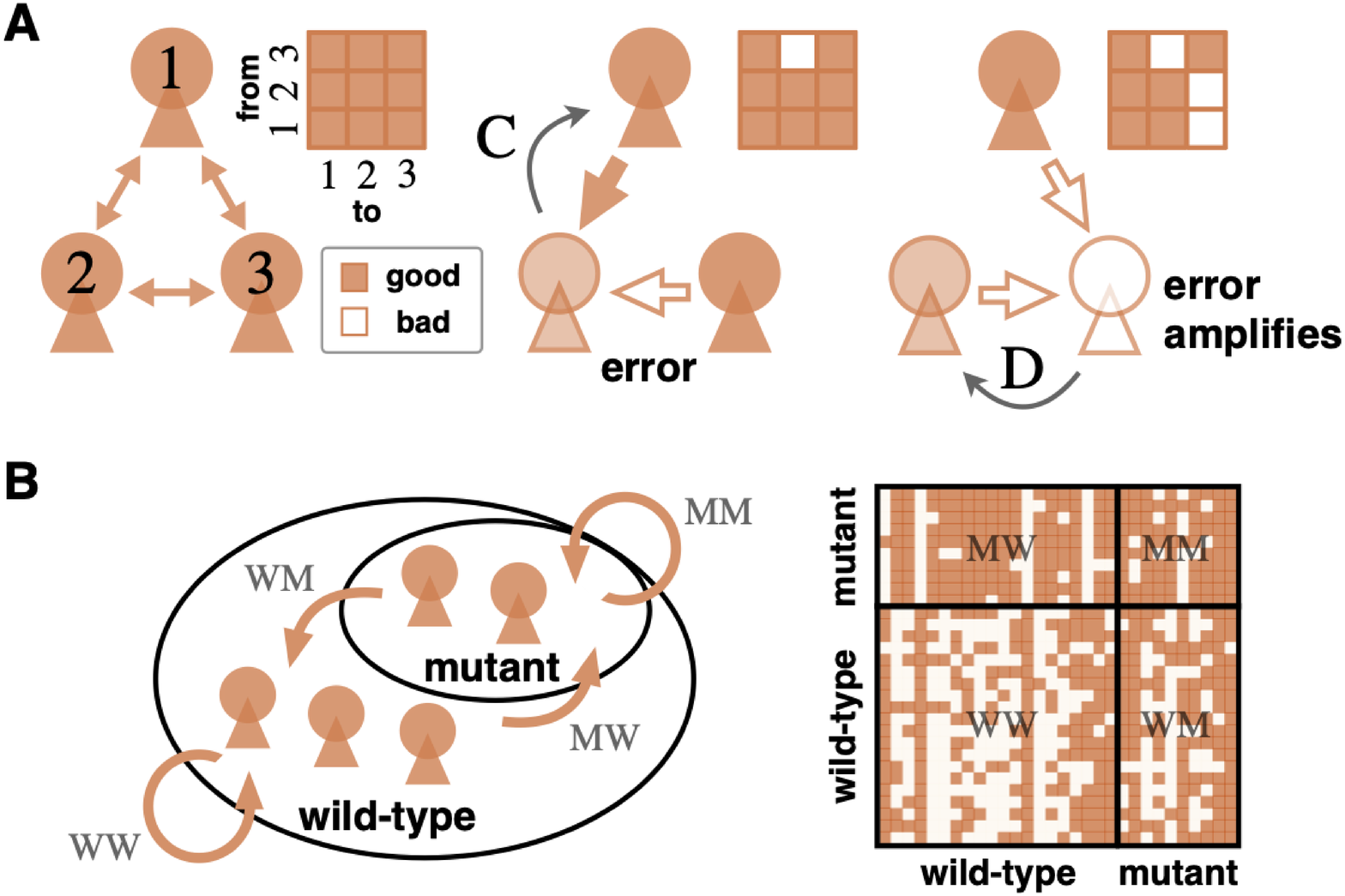}
\caption{{\bf A}. An illustration showing how an assessment error amplifies. In all the three panels, there are 1st, 2nd, and 3rd persons, and the $3\times 3$ image matrices and straight arrows indicate the reputations among them. In the left panel, good reputations are assigned among all of them, hence they achieve cooperation. In the center, the 2nd person cooperates with the 1st person, but the 3rd person erroneously assigns a bad reputation to the 2nd person. In the right, the 3rd person defects with the 2nd person based on its bad reputation in the eyes of the 3rd person, but because the 1st and 2nd persons believe that the 2nd person is good, they assign bad reputations to the 3rd person. {\bf B}. An illustration showing the complexity of the image matrix. The left panel shows that wild-types and mutants are mixed in the population and that four kinds of reputations exist; $\ww$ from wild-type to wild-type, $\wm$ from mutant to wild-type, $\mw$ from wild-type to mutant, and $\mm$ from mutant to mutant. The right panel shows that the image matrix is decomposed into the corresponding four components, each of which has a different reputation structure.
}
\label{F01}
\end{figure}

Most of these studies of private assessment have been performed by computer simulations~\cite{yamamoto2017norm, hilbe2018indirect}. This is because two-dimensional information of who assigns a reputation to whom (its matrix representation is called ``image matrix''~\cite{uchida2010effect, sigmund2012moral, uchida2013effect, oishi2013group}) becomes too complex to analyze. For example, its possible transition is illustrated in Fig.~\ref{F01}-A, where a single assessment error can be amplified with time, leading to a mosaic structure in the image matrix. An evolutionary analysis between wild-type and mutant makes the image matrix further complex because the image matrix now includes four compartments based on different rules of reputation assignment adopted by wild-type and mutant individuals (Fig.~\ref{F01}-B). In spite of these difficulties, here we report that we have successfully developed an analytical machinery to study the image matrix by applying a technique previously developed by the authors~\cite{fujimoto2022reputation}. This enables us to make a general prediction of when cooperation is sustained under noisy and private assessment over the full parameter region. 

In the following, we will first introduce the setting of indirect reciprocity under noisy and private assessment and explain a method to analytically calculate the expected payoff of each individual through analyzing a complex image matrix. Then, we will discuss which strategy can be an evolutionarily stable strategy (ESS)~\cite{smith1973logic, smith1982evolution} under which condition, and provide intuitive reasons for the result. To our knowledge, this is the first systematic study that has analytically investigated evolutionary stability of strategies in indirect reciprocity under noisy and private assessment.

\section*{Model}
We consider a model of indirect reciprocity in a well-mixed population of size $N$. We assume that, in every step, a binary reputation is assigned independently from everyone to everyone, either good or bad, which is summarized by image matrix $\{\beta_{ji}\}$, where $\beta_{ji}=1$ (resp. $\beta_{ji}=0$) if individual $i$ assigns a good (resp. bad) reputation to individual $j$. The model proceeds as follows. First, a donor and a recipient are randomly chosen from this population. Next, the donor takes its action, cooperation or defection, to the recipient. When the donor cooperates, the donor incurs a cost $c(>0)$ but gives a benefit $b(>c)$ to the recipient instead. On the other hand, when the donor defects, no change occurs in the payoff of the donor or the recipient. Here, a rule that specifies how the donor chooses its action is called ``action rule''. Throughout this paper, we assume that all the individuals adopt the ``discriminator'' action rule~\cite{nowak1998evolution,nowak1998dynamics}, with which they choose cooperation (resp. defection) to a good (resp. bad) recipient in their own eyes; that is, donor $i$ chooses cooperation toward recipient $j$ if $\beta_{ji}=1$, and chooses defection if $\beta_{ji}=0$. We assume that the donor unintentionally takes the opposite action to the intended one with probability $0\le e_1<1/2$ (action error). All the individuals in the population observe this social interaction between the donor and the recipient and independently update the reputation of the donor in their eyes.

A rule that specifies how each observer updates the reputations of the others is called its ``social norm''. In models of public reputation, it has often been assumed that all the individuals in the population adopt the same social norm~\cite{ohtsuki2004should,ohtsuki2007global,ohtsuki2015reputation} (but see \cite{uchida2010effect}), otherwise, they cannot share the reputation of the same individual. Because we consider a model of private reputation here, however, we instead assume that individuals can adopt different social norms. This study deals with a situation where each observer (say, $k$) refers to (i) whether the donor (say, $i$) cooperates (C) or defects (D) (first-order information) and (ii) whether the recipient (say, $j$) is good (G) or bad (B) in the eyes of the observer (second-order information, represented by $\beta_{jk}$) when this observer updates the reputation of the donor in the eyes of the observer, denoted by $\beta_{ik}$. Such social norms are called ``second-order'' social norms~\cite{nowak2005evolution,sigmund2010calculus,santos2018social1,santos2021complexity}.  An observer who adopts a second-order social norm can face four different cases, denoted by GC(``toward a Good recipient the donor Cooperates''), BC(``toward a Bad recipient the donor Cooperates''), GD(``toward a Good recipient the donor Defects''), and BD(``toward a Bad recipient the donor Defects''), respectively, and in each case, the observer assigns either a good (G) or bad (B) reputation to the donor. Thus, a social norm is represented by a four-letter string. For example, $\mathrm{GBBG}$ is the social norm that assigns to the donor a good reputation in GC- and BD-cases, and a bad reputation in BC- and GD-cases. There are $2^4=16$ such social norms in total, and we lexicographically order them with the rule that G comes first and B comes second, and number them from $S_{01}$ to $S_{16}$. Table \ref{tab:16norms} shows a full list of 16 social norms studied here. When updating the reputation, each observer independently commits an assessment error with probability $0<e_2<1/2$, in which case he/she accidentally assigns the opposite reputation to the intended one to the donor. 

\begin{table}%[tbhp]
\centering
\caption{All 16 second-order social norms in this study}
\begin{tabular}{cl|cccc}
    \multicolumn{2}{c|}{Social norm} & GC & BC & GD & BD \\
    \hline
    $S_{01}$ & (ALLG) & G & G & G & G\\
    $S_{02}$ & & G & G & G & B\\
    $S_{03}$ &(SS; Simple Standing) & G & G & B & G\\
    $S_{04}$ &(SC; Scoring) & G & G & B & B\\
    $S_{05}$ & & G & B & G & G\\
    $S_{06}$ & & G & B & G & B\\
    $S_{07}$ &(SJ; Stern Judging) & G & B & B & G\\
    $S_{08}$ &(SH; Shunning) & G & B & B & B\\
    $S_{09}$ & & B & G & G & G\\
    $S_{10}$ & & B & G & G & B\\
    $S_{11}$ & & B & G & B & G\\
    $S_{12}$ & & B & G & B & B\\
    $S_{13}$ & & B & B & G & G\\
    $S_{14}$ & & B & B & G & B\\
    $S_{15}$ & & B & B & B & G\\
    $S_{16}$ &(ALLB) & B & B & B & B
\end{tabular}
\label{tab:16norms}
\end{table}

Several norms are especially important in previous studies, so we explain them below. We call $S_{01}$ ALLG and call $S_{16}$ ALLB because these norms unconditionally assign good or bad reputations. Next, $S_{03}$, $S_{04}$, $S_{07}$, and $S_{08}$ belong to G$\ast$B$\ast$ family. These norms share the same feature that they regard cooperation toward a good recipient as good, and defection toward a good recipient as bad. They only differ when the recipient is bad in the observer's eyes. First, $S_{04}$ is called Scoring (SC), which regards cooperation toward a bad recipient as good and defection toward a bad recipient as bad, and therefore reputation assignment is independent of whether the recipient is good or bad in the observer's eyes (thus, categorized as a first-order norm). Next, $S_{07}$ is called Stern Judging (SJ), which regards cooperation toward a bad recipient as bad and defection toward a bad recipient as good, as opposed to SC. Third, $S_{03}$ is called Simple Standing (SS) and it regards any action toward a bad recipient as good, and therefore it is the most generous norm in this family. Finally, $S_{08}$ is called  Shunning (SH) and it regards any action toward a bad recipient as bad, and therefore it is the most intolerant one. Notably, SJ and SS are the two second-order norms that are included in the ``leading eight'' norms~\cite{ohtsuki2004should}, which are norms that can successfully maintain cooperation under the noisy and public reputation that are found in the search within third-order norms. In particular, SJ has long been considered promising because it is evolutionarily successful~\cite{pacheco2006stern} and because it sustains a very high level of cooperation despite its simplicity~\cite{santos2018social1, santos2021complexity}. SJ always suggests only one correct action to keep you good; it recommends cooperation toward good individuals and defection toward bad ones, and failure to follow this rule leads to a bad reputation. Under the noisy public reputation, SH cannot achieve full cooperation against itself but can prevent the invasion of ALLB (see SI for detailed calculation).

Under these settings, the strategy of an individual is its social norm. For this reason, we use ``strategy'' and ``(social) norm'' interchangeably in the following. We ask which strategy is evolutionarily stable. To this end, we study invasibility of a mutant strategy against a wild-type one. A strategy is ESS if it is not invaded by any other 15 mutant strategies. To derive their payoffs, we need to analyze the image matrix, which we shall perform below.

\section*{Analysis of reputation structure}
Let us consider a situation where individuals with mutant norm ${\rm M}$  invade the population of wild-type norm ${\rm W} (\neq {\rm M})$. Here, the proportion of mutants is given by $\delta$. By extending the Fujimoto \& Ohtsuki's method~\cite{fujimoto2022reputation} we can describe the image matrix by two probability distributions. Specifically, take a focal individual whose norm is $A \in\{{\rm W},{\rm M}\}$, and let $p_{AA'}$ (hereafter called ``goodness'') be the proportion of individuals among norm $A'$ users who assign a good reputation to the focal individual, for $A'\in\{{\rm W},{\rm M}\}$. Thus, a wild-type individual is characterized by a pair of goodnesses, $(p_{\ww},p_{\wm})$, and we represent its distribution over all wild-type individuals by $\Phi_{\rm W}(p_{\ww},p_{\wm})$. In the same way, a mutant is characterized by a pair of goodnesses, $(p_{\mw},p_{\mm})$, and $\Phi_{\rm M}(p_{\mw},p_{\mm})$ represents its distribution over all mutants. In SI, we derive the dynamics of $\Phi_{\rm W}$ and $\Phi_{\rm M}$ by formulating a stochastic transition of the donor's goodnesses under the assumption of $N \gg 1$ (the population is large), $\delta \ll 1$ (mutants are rare), and $N\delta \gg 1$ (yet the number of mutants is sufficiently large). Then, we derive the equilibrium distributions, $\Phi_{\rm W}^{*}$ and $\Phi_{\rm M}^{*}$. These equilibrium distributions give expected payoffs of wild-types and mutants, which enable us to study the invasibility condition of mutants to wild-types (see SI again).

% Figure 2
\begin{figure}[!ht]%[tbhp]
\centering
\includegraphics[width=0.6\linewidth]{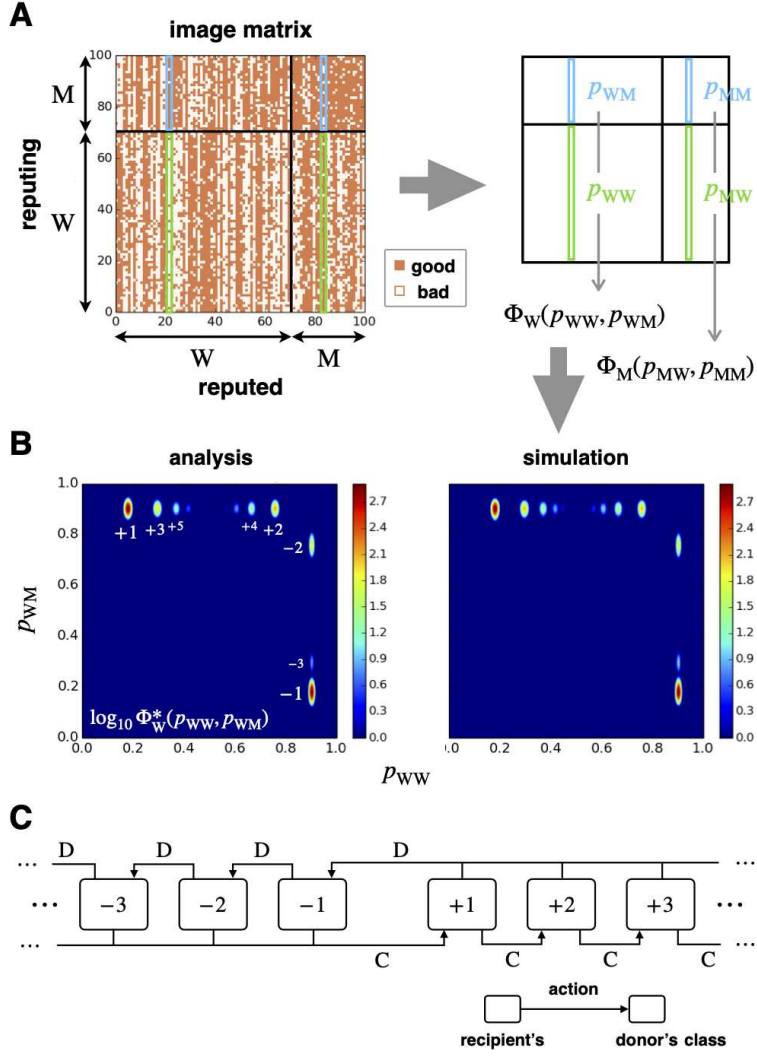}
\caption{Illustrations of our method to analyze reputation structure. {\bf A}. In the left panel, an example of image matrix is shown. We analyze this image matrix divided into four parts; whether the reputing side is wild-type (W) or mutant (M) and whether the reputed side is W or M. In the right panel, a pair of goodnesses of each individual from wild-types (colored green) and mutants (blue) are extracted from the image matrix. Because the pair of goodnesses correlate with each other, we consider the joint probability distribution of them, denoted by $\Phi_{\rm W}$ and $\Phi_{\rm M}$. {\bf B}. We analytically calculated this joint probability distribution. One can see that the analytical estimation (the left panel) well fits the simulation (right). In both the panels, we assume $({\rm W},{\rm M})=(S_{09},S_{03})$, $N=5000$, $\delta=0.1$, and $(e_1,e_2)=(0,0.1)$. In the numerical simulation, we used $3000$ samples of image matrices from time $t=51,\cdots,3050$ (a random donor's goodness is updated $N$ times per unit time of $t$). On the other hand, in the theoretical analysis, we introduced the cutoff of $-100\le j\le +100$. Each number near the heat peaks indicates the class label $j$. {\bf C}. Rules for labeling individual classes. Each class corresponds to one Gaussian distribution. Each box (labelled by $j\in\mathbb{Z}\backslash\{0\}$) indicates a class. The destination of each arrow indicates the class that the donor moves to after taking cooperation (C) or defection (D) toward the recipient that belongs to the class that the arrow originates. For example, a donor that cooperated with class $j=-2$ recipient moves to class $j=+1$.}
\label{F02}
\end{figure}

We find that each of the two equilibrium distributions is well approximated by a weighted sum of two-dimensional Gaussian functions with zero covariance, where each Gaussian can be systematically labeled by a nonzero integer, $j\in\mathbb{Z}\backslash\{0\}$ (see an example in Fig.~\ref{F02}-B and the rule of labeling in Fig.~\ref{F02}-C). Hence the number of Gaussians that appear in the sum is infinitely but countably many. In some cases, however, these labels degenerate (i.e., two or more Gaussians are identical but they are given different labels) and the number of Gaussians can be finite. Weights to Gaussians decay exponentially as $j$ becomes large positive or large negative, so a truncation at some finite number of terms approximates well the infinite sum for numerical calculations.

\section*{ESS norms}
Based on the analysis of the image matrix above, we have studied pairwise invasibility for all the pairs of wild-type $\rm W$ and mutant $\rm M$. In the following, we set the action error rate as $e_1=0$, because this error, especially when it is small positive, does not have a qualitative impact on our results as far as we studied. Thus, the cost-benefit ratio $b/c$ and the assessment error rate $e_2$ are our environmental parameters.

We first find that the four strategies, $S_{06}$, $S_{07}$(SJ), $S_{10}$, and $S_{11}$, are completely indistinguishable, both as wild-types and as mutants. This is because these norms always give the goodness of $1/2$ to anyone in the population at equilibrium due to an accumulation of assessment errors and hence they appear to choose cooperation and defection in a random manner. In particular, they are neutral to each other. For these reasons, we will discuss only $S_{07}$(SJ) as a representative of them and exclude the other three in the following analysis.

\begin{figure*}
\centering
\includegraphics[width=0.8\linewidth]{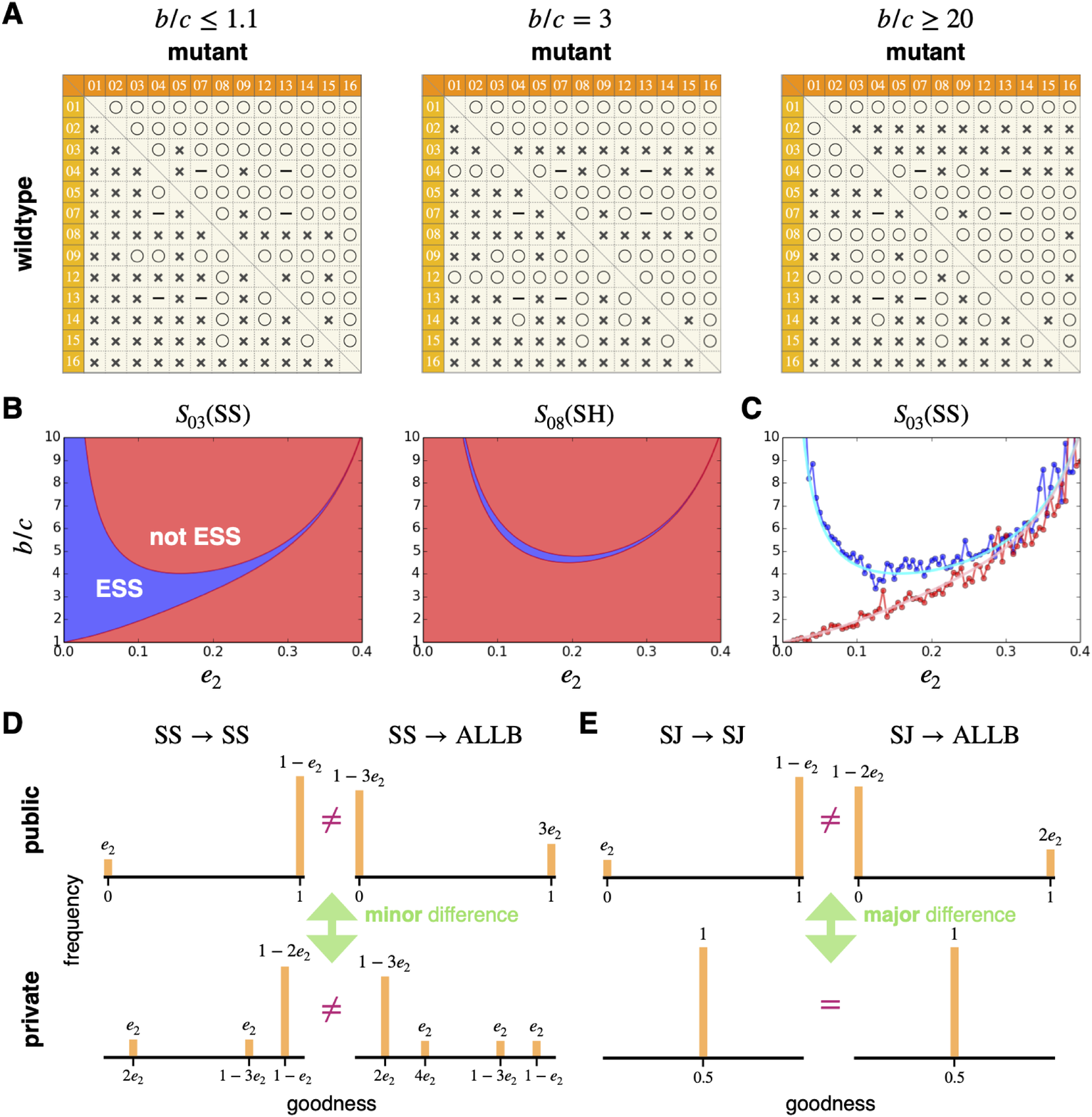}
\caption{Details of ESS analysis. {\bf A}. Invasibility between all pairs of the social norms. The left, center, and right panels respectively show the cases of $b/c\le1.1$, $=3$, and $\ge 20$. Numbers in rows (resp. columns) indicate the labels of wild-type (resp. mutant) norms. Each circle (resp. cross) mark indicates that the invasion by mutants is successful (resp. unsuccessful). Each bar mark indicates that the wild-type and mutant norms are neutral. All the panels are based on $e_2=0.1$ and $-10000\le j\le +10000$. {\bf B}. The ESS parameter region for norms $S_{03}$(SS) (left) and $S_{08}$(SH) (right). In each panel, the horizontal (resp. vertical) axis indicates $e_2$ (resp. $b/c$). The blue (resp. red) color indicates that the norm is ESS (resp. not ESS). {\bf C}. Comparison between analytical and numerical calculations of the ESS region of $S_{03}$(SS). The horizontal and vertical axis are the same as in {\bf B}. The cyan (resp. pink) line indicates the theoretical upper (resp. lower) bound (the same as the left panel in {\bf B}). Blue (resp. red) dots, connected by lines, indicate the numerical estimates of the upper (resp. lower) bound. Those estimates were calculated based on agent-based simulations of image matrix with $N=10000$, $\delta=0.03$. The average of $50$ samples from generations $51 \leq t \leq 100$ were used, except for in the calculation of the upper bound (blue dots) for $e_2<0.1$ where we instead used the average of $3000$ samples from $51 \leq t \leq 3050$ to reduce errors in estimation. {\bf D}. A comparison between public (top row) and private (bottom row) assessment cases of how wild-type SS individuals evaluate other SS individuals (left column) and how wild-type SS individuals evaluate mutant ALLB individuals (right column). In each panel, the horizontal and vertical axes indicate individual goodness and its frequency, respectively. Positions and heights of bars are correct only up to order $e_2$. We see that SS gives high goodness to most of the SS individuals (left column), that SS gives low goodness to most of the ALLB individuals (right column), and that the difference between the top and bottom rows is minor (in a scale of $O(e_2)$). Thus, SS is robust against the invasion by ALLB under both public and private assessment. {\bf E}. Similar comparison to {\bf D} was made for wild-type SJ and mutant ALLB. We see that SJ gives high goodness to most of the SS individuals under public assessment (top left), that SJ gives low goodness to most of the ALLB individuals (top right), but that SJ gives goodness of $1/2$ to both SJ (bottom left) and ALLB (bottom right) individuals under private assessment. Thus, SJ is robust against the invasion by ALLB under public assessment while it is not under private assessment.}
\label{F04}
\end{figure*}

Our exhaustive analysis demonstrates that only three norms, $S_{03}$(SS), $S_{08}$(SH), and $S_{16}$(ALLB) can be ESS, and all the others cannot. As shown in Fig.~\ref{F04}-A, ALLB is ESS independent of $b/c$ and $e_2$, because it is the norm that assigns a bad reputation to everyone, saves the own cost, and provides no benefit to others. On the other hand, SS and SH achieve ESS for some $b/c$ and $e_2$; there are upper and lower bounds of $b/c$ for them to be ESS, which depend on $e_2$. Below we will look at its details.

\section*{Conditions for ESS}
The ESS condition of $S_{03}$(SS) is shown in Fig.~\ref{F04}-B. When $b/c$ exceeds the upper bound, the norm is invaded by $S_{01}$(ALLG) (compare the right and center panels of Fig.~\ref{F04}-A). On the other hand, when $b/c$ falls below the lower bound, the norm is invaded by $S_{04}$(SC) (compare the left and center panels of Fig.~\ref{F04}-A). Fig.~\ref{F04}-C shows that these theoretical bounds are also supported by individual-based simulations. Notably, the smaller $e_2$ is, the wider the ESS region of $S_{03}$(SS) becomes.

The ESS region of $S_{08}$(SH) is quite narrow in comparison to that of $S_{03}$(SS), as seen in Fig.~\ref{F04}-B. In addition, when $b/c$ exceeds the upper bound or falls below the lower bound, the norm is invaded by $S_{04}$(SC) and $S_{16}$(ALLB), respectively. The range of $b/c$-ratios that make SH evolutionarily stable is the widest at an intermediate $e_2$ (about $0.1$).

In contrast to these results, we find that $S_{07}$(SJ), which is known to be a successful norm when reputation is public, is invaded by norms such as $S_{16}$(ALLB) and $S_{08}$(SH) independent of the value of $b/c$ (and also independent of $e_2$), and therefore that it is never an ESS. This is summarized in Fig.~\ref{F04}-A.

To summarize, SS, SH, and SJ are all the ESS norms under the public assessment, but whether they remain ESS in the private assessment critically differs. This difference is clearly understood by focusing on how the reputation structure they give differs between the public and private reputation cases under a sufficiently small but positive assessment error rate, $e_2\ll 1$. Let us consider below, for example, whether each norm can prevent the invasion of ALLB, a potential invader norm.

{\bf Success of Simple Standing:} The reputation structure that $S_{03}$(SS) gives differs little between the public and private reputation cases (see Fig.~\ref{F04}-D). Under the public reputation (see SI for the calculation), SS assigns good reputations to SS themselves (represented by the bar at goodness $=1$ in the top-left panel in Fig.~\ref{F04}-D), while bad reputations to ALLB (represented by the bar at goodness $=0$ in the top-right panel in Fig.~\ref{F04}-D). Thus, SS distinguishes between SS itself and the invader ALLB and prevents the invasion of ALLB. Even under the private reputation, SS still assigns good reputations to SS themselves (see the bottom-left panel in Fig.~\ref{F04}; high goodness of $1-e_2$ are given to the fraction $1-2e_2$ of SS individuals, for example), and assigns bad reputations to ALLB (see the bottom-right panel in Fig.~\ref{F04}; low goodness of $2e_2$ are given to the fraction $1-3e_2$ of ALLB individuals, for example). Thus, the distinction between SS and ALLB is maintained. For that reason, SS succeeds in achieving ESS even under the private assessment. The cooperation rate at this ESS is as high as $1-2e_2$ for small $e_2$, so it entails nearly perfect cooperation.

{\bf Failure of Stern Judging:} Contrary to SS, the reputation structure that $S_{07}$(SJ) gives extremely differs between the public and private reputation cases (see Fig.~\ref{F04}-E). Under the public reputation (see SI for the calculation), wild-type SJ gives high goodness to other SJ (top-left in Fig.~\ref{F04}-E) and wild-type SJ gives low goodness to ALLB (top-right in Fig.~\ref{F04}-E). Thus, SJ prevents the invasion of ALLB. Under the private assessment, however, SJ gives goodness of 1/2 to other SJ individuals (bottom-left in Fig.~\ref{F04}-E)~\cite{uchida2013effect} while SJ gives goodness of 1/2 to ALLB individuals as well (bottom-right in Fig.~\ref{F04}-E). Thus, the distinction between SJ and ALLB is lost. This is why SJ fails to be ESS under the private assessment.

{\bf Shunning can be ESS, but the level of cooperation is low:} We can understand why $S_{08}$(SH) achieves ESS only in a narrow region under private reputation (see SI for the detailed calculation and see Fig.~S3 for the illustration for easy interpretation). Under the public reputation, SH gives good reputations to the half of other SH and bad reputations to the other half (top-left in Fig.~S3) while SH gives bad reputations to almost all ALLB (top-right in Fig.~S3). Thus, SH prevents the invasion of ALLB. Under private reputation, on the other hand, SH gives low goodness to both SH and ALLB (bottom-left and bottom-right in Fig.~S3). Here, however, SH has a slightly better chance to receive good reputations than ALLB, in the order of $e_2^2$. This explains why SH prevents the invasion from ALLB only in a narrow region and also explains why its ESS condition becomes more strict for a smaller assessment error rate, $e_2$. The cooperation rate at a realized ESS is as low as $e_2$ for small $e_2$, so we conclude that $S_{08}$(SH) does not contribute to cooperation.

\section*{Discussion}
This study considered indirect reciprocity under noisy and private assessment. We focused on goodness of individual (i.e., what proportion of individuals gives the individual good reputations) between different norms and developed an analytical method to calculate the distribution of goodness at equilibrium. Using this methodology we studied whether a mutant norm succeeds in the invasion into a wild-type norm. Although both $S_{03}$(SS) and $S_{07}$(SJ) can be ESS under public reputation, we found that their evolutionary stability is totally different under private assessment. In particular, we found that $S_{03}$(SS) remains to be ESS under private assessment if the assessment error rate is small, while $S_{07}$(SJ) cannot be ESS no matter how small the error rate is. 

The reason for this difference between $S_{03}$(SS) and $S_{07}$(SJ) comes from the difference in the complexity of these two norms. In the world of private assessment, errors in assessment accumulate independently among observers, which is a potential source of collapse of cooperation in the population. However, since $S_{03}$(SS) regards a cooperating donor as good no matter whether the recipient is good or bad, a discrepancy in the opinion toward the recipient between two different observers does not produce further discrepancy; those two observers can agree that such a cooperating donor is good. In contrast, $S_{07}$(SJ) is more complex than $S_{03}$(SS) and recipient's reputation is always decisive information (see Table \ref{tab:16norms}), so this complexity becomes an obstacle for correcting discrepancy between observers.  

Hilbe et al.~\cite{hilbe2018indirect} studied by computer simulations whether the leading eight norms can sustain cooperation under the noisy and private assessment. They concluded that $S_{03}$(SS) (referred to as ``L3'' in their paper) and $S_{07}$(SJ) (``L6'') fail to achieve cooperation, which is contrary to our result. This difference is because we studied evolutionary stability in a deterministic model, while they studied fixation probability in a stochastic model. Because those two criteria are different, drawing a general conclusion is difficult, but the significance of our study lies in that we have shown that cooperation can be evolutionarily sustained even under private assessment.

A future direction of this study would be to examine ESS conditions of social norms when some of the assumptions are changed. For example, we have assumed second-order norms, in which individuals refer to a donor's action (first-order information) and a recipient's reputation (second-order one) when they update the donor's reputation. However, humans may use more complex norms than the second-order ones. Studying the effect of higher-order information ~\cite{sugden1986economics, sasaki2017evolution, santos2018social1, santos2021complexity}, such as the previous reputation of the donor (third-order information), would further deepen our understanding. We have also assumed that all individuals simultaneously update their opinions toward the same donor. However, in a real society, the number of people who can observe a single person's behavior is limited. Thus, the effect of asynchronous updates of reputations is worth studying. Last but not least, we have implicitly assumed that game interactions last sufficiently long so that we can use equilibrium distributions of goodness for calculating payoffs (i.e. discount factor is $1$). However, the effect of initial reputation cannot necessarily be ignored in some cases.

In conclusion, we have demonstrated that cooperation can be evolutionarily stable even under the noisy and private assessment. Specifically, we have shown that Stern Judging, which is one of the most leading norms under public reputation, cannot distinguish between cooperators and defectors under private assessment and thus fails to achieve ESS. On the other hand, we have revealed that Simple Standing can be stable in a wide range of parameters. Based on these results, we predict that Simple Standing should play a key role in sustaining cooperation by indirect reciprocity under noisy private assessment. These findings provide a rigid theoretical basis for understanding human cooperation and pave the way for future studies in biology, psychology, sociology, and economics.

\section*{Acknowledgement}
Y.F. acknowledges the support by JSPS KAKENHI Grant Number JP21J01393. H.O. acknowledges the support by JSPS KAKENHI Grant Number JP19H04431.

%\bibliographystyle{unsrt}
%\bibliography{biblio.bib}

\clearpage

\setcounter{section}{0}
\setcounter{figure}{0}
\renewcommand{\figurename}{FIG. S}
\renewcommand{\thesection}{S\arabic{section}}

\begin{center}
{\LARGE {\bf Supplementary Material}}
\end{center}

\small

\section{Calculation of joint distribution of goodnesses}
This section proposes an analytical method to obtain the reputation structure under indirect reciprocity. We assume a situation where rare mutants with norm ${\rm M}$ of ratio $\delta$ invade other wild-types with norm ${\rm W}$ of ratio $1-\delta$. We denote the population ratio of norm $A\in\{{\rm W},{\rm M}\}$ as $\rho_A$; $\rho_A=1-\delta$ when $A={\rm W}$, while $\rho_A=\delta$ when $A={\rm M}$. To characterize the reputation structure, we define $p_{iA}$ as a proportion of individuals of norm $A$ who assign good reputations to individual $i$. We call $p_{iA}$ a goodness of individual $i$ from norm $A\in\{{\rm W},{\rm M}\}$.

In the following, let us consider a stochastic transition of $p_{iA}$ in each round. In a single round, a recipient and a donor are chosen and labeled as $i_{\rm R}$ and $i_{\rm D}$, respectively. In this round, $p_{i_{\rm D}A_{\rm O}}$, i.e., the goodness of donor from norm $A_{\rm O}$, changes into the next goodness $p_{i_{\rm D}A_{\rm O}}'$ for all $A_{\rm O}\in\{{\rm W},{\rm M}\}$. Below, we formulate the stochastic change separately for cases that the donor chooses to cooperate or defect.

{\bf C-map case:} First, we consider a case that the donor cooperates with the recipient, occurring with a probability of
\begin{align}
    h(p_{i_{\rm R}A_{\rm D}}):=p_{i_{\rm R}A_{\rm D}}(1-e_1)+(1-p_{i_{\rm R}A_{\rm D}})e_1.
    \label{Cprob}
\end{align}
In this case, $N\rho_{A_{\rm O}}p_{i_{\rm D}A_{\rm O}}'$, i.e., the number of observers with norm $A_{\rm O}$ who give good reputations to the donor in the next round, follows a probability distribution of
\begin{align}
    &N\rho_{A_{\rm O}}p_{i_{\rm D}A_{\rm O}}'\sim N_1+N_2,\\
    \label{N'_C}
    &N_1\sim {\cal B}(N\rho_{A_{\rm O}}p_{i_{\rm R}A_{\rm O}},a^{\rm GC}_{A_{\rm O}}),\\
    &N_2\sim {\cal B}(N\rho_{A_{\rm O}}(1-p_{i_{\rm R}A_{\rm O}}),a^{\rm BC}_{A_{\rm O}}).
\end{align}
Here, ${\cal B}(n,a)$ denotes a binomial distribution with success probability $a$ and trial number $n$. In addition, $a^{XY}_{A_{\rm O}}$ denotes the probability that an observer with norm $A_{\rm O}$ who evaluates the recipient as $X\in\{{\rm G},{\rm B}\}$ newly gives a good reputation to the donor whose action is $Y\in\{{\rm C},{\rm D}\}$. $a^{\rm GC}_{A_{\rm O}}$, $a^{\rm BC}_{A_{\rm O}}$, $a^{\rm GD}_{A_{\rm O}}$, and $a^{\rm BD}_{A_{\rm O}}$ are obtained by converting corresponding $\rm G$ and $\rm B$ pivots into $1-e_2$ and $e_2$ in Table~1 of the main manuscript. Instead of \eqref{N'_C}, we use a shorthand notation;
\begin{align}
    N\rho_{A_{\rm O}}p_{i_{\rm D}A_{\rm O}}'\sim {\cal B}(N\rho_{A_{\rm O}}p_{i_{\rm R}A_{\rm O}},a^{\rm GC}_{A_{\rm O}})+{\cal B}(N\rho_{A_{\rm O}}(1-p_{i_{\rm R}A_{\rm O}}),a^{\rm BC}_{A_{\rm O}}).
\end{align}
Because $N\rho_{A_{\rm O}}$ is sufficiently large, the mean and variance of $p_{i_{\rm D}A_{\rm O}}'$ are given by
\begin{align}
        \textrm{E}[p_{i_{\rm D}A_{\rm O}}']&=p_{i_{\rm R}A_{\rm O}}(\underbrace{a^{\rm GC}_{A_{\rm O}}-a^{\rm BC}_{A_{\rm O}}}_{=:\Delta f_{A_{\rm O}}^{\rm C}})+a^{\rm BC}_{A_{\rm O}}\quad (=: f_{A_O}^{\rm C}(p_{i_{\rm R}A_{\rm O}})),
        \label{f_C} \\
        \textrm{Var}[p_{i_{\rm D}A_{\rm O}}']&=\frac{p_{i_{\rm R}A_{\rm O}}a^{\rm GC}_{A_{\rm O}}(1-a^{\rm GC}_{A_{\rm O}})+(1-p_{i_{\rm R}A_{\rm O}})a^{\rm BC}_{A_{\rm O}}(1-a^{\rm BC}_{A_{\rm O}})}{N\rho_{A_{\rm O}}}=\frac{e_2(1-e_2)}{N\rho_{A_{\rm O}}}\quad (=: \rho_{A_{\rm O}}^{-1}s^2).
        \label{s2}
\end{align}
In \eqref{f_C}, $f_{A_{\rm O}}^{\rm C}$ represents a map from the recipient's goodness in the present round to the donor's goodness in the next round. Because this map is applied only when the donor cooperates, we call it ``C-map''.

{\bf D-map case:} On the other hand, we consider a case that the donor defects with the recipient, occurring with a probability of
\begin{align}
    1-h(p_{i_{\rm R}A_{\rm D}})=(1-p_{i_{\rm R}A_{\rm D}})(1-e_1)+p_{i_{\rm R}A_{\rm D}}e_1.
    \label{Dprob}
\end{align}
In this case, $N\rho_{A_{\rm O}}p_{i_{\rm D}A_{\rm O}}'$ follows a probability distribution of
\begin{align}
    &N\rho_{A_{\rm O}}p_{i_{\rm D}A_{\rm O}}'\sim
        {\cal B}(N\rho_{A_{\rm O}}p_{i_{\rm R}A_{\rm O}},a^{\rm GD}_{A_{\rm O}})+{\cal B}(N\rho_{A_{\rm O}}(1-p_{i_{\rm R}A_{\rm O}}),a^{\rm BD}_{A_{\rm O}}).
    \label{N'_D}
\end{align}
From this equation, the mean and variance of $p_{i_{\rm D}A_{\rm O}}'$ are given by
\begin{align}
        \textrm{E}[p_{i_{\rm D}A_{\rm O}}']&=p_{i_{\rm R}A_{\rm O}}(\underbrace{a^{\rm GD}_{A_{\rm O}}-a^{\rm BD}_{A_{\rm O}}}_{=:\Delta f_{A_{\rm O}}^{\rm D}})+a^{\rm BD}_{A_{\rm O}}\quad (=: f_{A_{\rm O}}^{\rm D}(p_{i_{\rm R}A_{\rm O}})),
        \label{f_D} \\
        \textrm{Var}[p_{i_{\rm D}A_{\rm O}}']&=\frac{p_{i_{\rm R}A_{\rm O}}a^{\rm GD}_{A_{\rm O}}(1-a^{\rm GD}_{A_{\rm O}})+(1-p_{i_{\rm R}A_{\rm O}})a^{\rm BD}_{A_{\rm O}}(1-a^{\rm BD}_{A_{\rm O}})}{N\rho_{A_{\rm O}}}=\frac{e_2(1-e_2)}{N\rho_{A_{\rm O}}}\quad (=: \rho_{A_{\rm O}}^{-1}s^2).
\end{align}
Because the map $f_{A_{\rm O}}^{\rm D}$ is applied when the donor defects, we call it D-map in the same way as C-map.

\begin{figure}
    \centering
    \includegraphics[width=0.85\hsize]{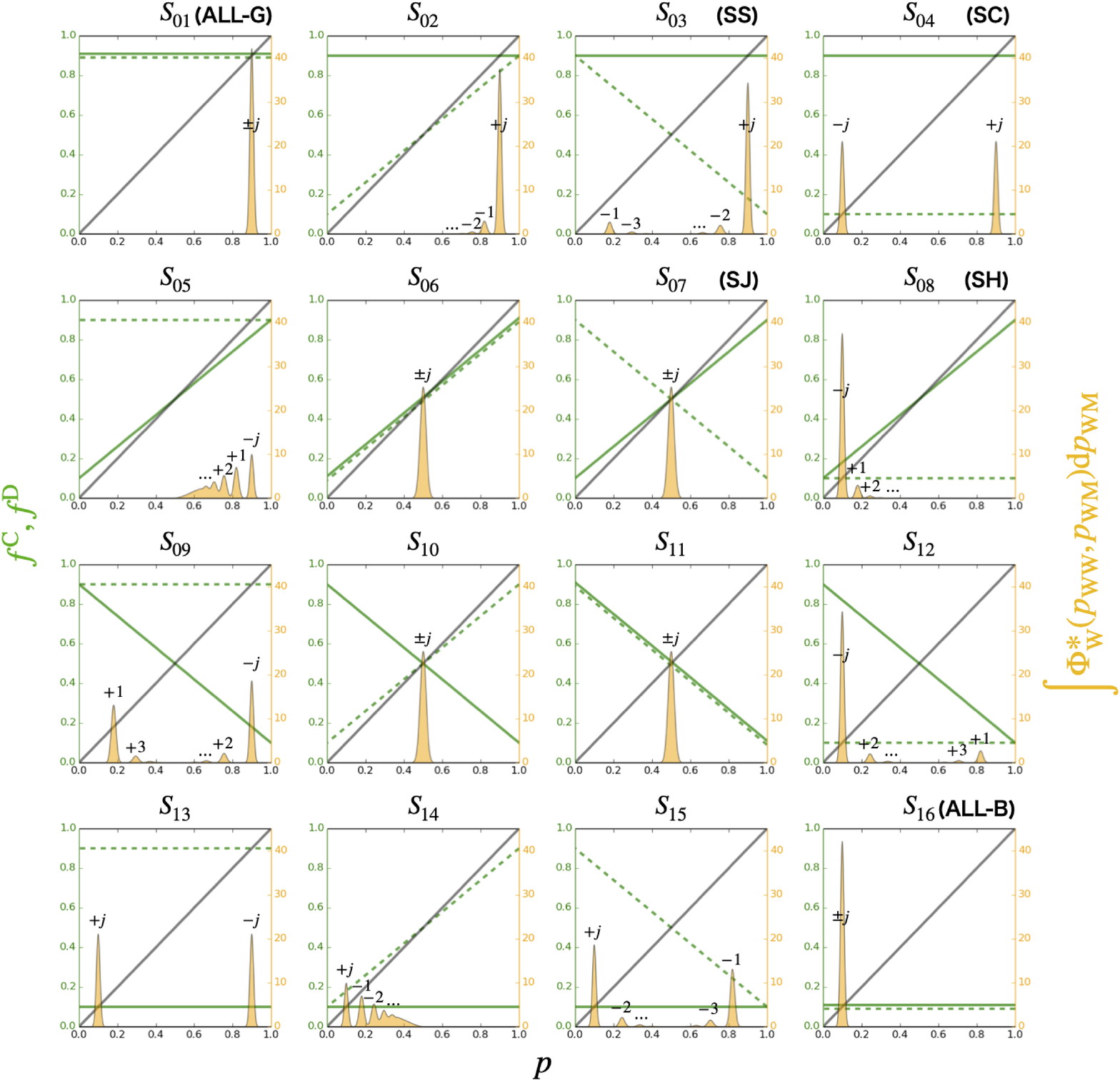}
    \caption{Materials for the reputation structure for all the second-order norms ${\rm W}=S_k$. Green solid (resp. broken) lines indicate C-map $f_{\rm W}^{\rm C}$ (resp. D-map $f_{\rm W}^{\rm D}$) of the norm. Gray lines indicate the identity map, which shows the fixed points of the C-map and D-map as the crossing points with these maps. The orange distribution shows the probability density function of goodnesses $p_{\ww}$ when ${\rm W}=S_k$. All the panels are output under $N=2000$, $\delta=0$, and $(e_1,e_2)=(0,0.1)$. Numbers over each peak indicate $j$.}
    \label{FS01}
\end{figure}

The above C-map $f_{S_k}^{\rm C}$ and D-map $f_{S_k}^{\rm D}$ are illustrated in Fig.~\ref{FS01} for all $S_k\in{\cal S}$.
% Figure S01

\clearpage

\section{Time evolution of reputation structure}
Because the population of wild-types and mutants are sufficiently large, we can continualize the distribution of individual goodness $p_{iA}$ with separating into the cases that the norm of individual $i$ is ${\rm W}$ or ${\rm M}$. In the following, $p_{AA'}$ denotes a continualized goodness of an individual with norm $A$ in the eyes of individuals with norm $A'$. Let us consider a time change of the distribution of $p_{AA'}$. As shown above, however, we should keep in mind that $p_{A{\rm W}}$ and $p_{A{\rm M}}$ are simultaneously changed by the C-map or D-map. Thus, we consider dynamics of $\Phi_{A}(p_{A{\rm W}},p_{A{\rm M}})$, a joint probability distribution of $p_{A{\rm W}}$ and $p_{A{\rm M}}$. Note that a norm of the chosen recipient is ${\rm M}$ only with a probability of $\delta$, which contributes the dynamics of $\Phi_{A}$ only in a scale of $O(\delta)$. By ignoring this scale of $O(\delta)$, the dynamics of $\Phi_{A}$ is given by
\begin{align}
    \frac{\mathrm{d}}{\mathrm{d}t}\Phi_{A}(p_{A{\rm W}},p_{A{\rm M}})=-\Phi_{A}(p_{A{\rm W}},p_{A{\rm M}})+\int_0^1\int_0^1&\{h(p'_{{\rm W}A})g(p_{A{\rm W}};f_{\rm W}^{\rm C}(p'_{\rm WW}),\rho_{\rm W}^{-1}s^2)g(p_{A{\rm M}};f_{\rm M}^{\rm C}(p'_{\rm WM}),\rho_{\rm M}^{-1}s^2)
    \nonumber\\
    &+(1-h(p'_{{\rm W}A}))g(p_{A{\rm W}};f_{\rm W}^{\rm D}(p'_{\rm WW}),\rho_{\rm W}^{-1}s^2)g(p_{A{\rm M}};f_{\rm M}^{\rm D}(p'_{\rm WM}),\rho_{\rm M}^{-1}s^2)\}
    \nonumber\\
    &\times\Phi_{\rm W}(p'_{\rm WW},p'_{\rm WM})\mathrm{d}p'_{\rm WW}\mathrm{d}p'_{\rm WM}.
    \label{int-dif}
\end{align}
Here, $g(p;\mu,\sigma^2)$ denotes a Gaussian function with the mean $\mu$ and variance $\sigma^2$ as
\begin{align}
    g(p;\mu,\sigma^2):=\frac{1}{\sqrt{2\pi\sigma^2}}\exp\left(-\frac{(p-\mu)^2}{2\sigma^2}\right).
\end{align}
Equation~(\ref{int-dif}) explains an update of the donor's goodness per time. The first (resp. second) term in the right side represents decrements (increments) by updating goodnesses. In detail, $\Phi_{\rm W}(p'_{\ww},p'_{\wm})$ in the second term shows the density that the recipient's goodness is $p'_{\ww}$ (resp. $p'_{\wm}$) in the eyes of wild-types (resp. mutants). $h(p'_{{\rm W}A})$ shows the probability that the donor cooperates, and the donor's goodnesses after the update in the eyes of observers with norm ${\rm W}$ and ${\rm M}$ are described by $g(p_{A{\rm W}};f_{\rm W}^{\rm C}(p'_{\ww}),\rho_{\rm W}^{-1}s^2)$ and $g(p_{A{\rm M}};f_{\rm M}^{\rm C}(p'_{\wm}),\rho_{\rm M}^{-1}s^2)$, respectively. A similar explanation holds when the donor chooses to defect.

The equilibrium state of \eqref{int-dif}, i.e., $\Phi_{A}^*$, satisfies
\begin{align}
    \Phi_{A}^*(p_{A{\rm W}},p_{A{\rm M}})=\int_0^1\int_0^1&\{h(p'_{{\rm W}A})g(p_{A{\rm W}};f_{\rm W}^{\rm C}(p'_{\rm WW}),\rho_{\rm W}^{-1}s^2)g(p_{A{\rm M}};f_{\rm M}^{\rm C}(p'_{\rm WM}),\rho_{\rm M}^{-1}s^2)
    \nonumber\\
    &+(1-h(p'_{{\rm W}A}))g(p_{A{\rm W}};f_{\rm W}^{\rm D}(p'_{\rm WW}),\rho_{\rm W}^{-1}s^2)g(p_{A{\rm M}};f_{\rm M}^{\rm D}(p'_{\rm WM}),\rho_{\rm M}^{-1}s^2)\}
    \nonumber\\
    &\times\Phi_{\rm W}^*(p'_{\rm WW},p'_{\rm WM})\mathrm{d}p'_{\rm WW}\mathrm{d}p'_{\rm WM}.
    \label{equilibrium}
\end{align}
To solve this equation, we assume that the equilibrium state can be described by a summation of two-dimensional Gaussian functions without correlation as
\begin{align}
    \Phi_{A}^*(p_{A{\rm W}},p_{A{\rm M}})=\sum_jq_{Aj}g(p_{A{\rm W}};\mu_{A{\rm W}j},\rho_{\rm W}^{-1}\sigma^2_{A{\rm W}j})g(p_{A{\rm M}};\mu_{A{\rm M}j},\rho_{\rm M}^{-1}\sigma^2_{A{\rm M}j}).
    \label{sum_gaussian}
\end{align}
The assumption of Gaussian is justified by the above transition process of the donor's goodness, where the goodness is virtually determined only by the mean and variance in a sufficiently large population. No correlation is assumed because the variance is given independently by observers with different norms.

We now derive equations which the equilibrium state satisfies for each norm $A\in\{{\rm W},{\rm M}\}$. First, substituting \eqref{sum_gaussian} into \eqref{equilibrium} for $A={\rm W}$, we obtain
\begin{align}
    &\sum_jq_{{\rm W}j}g(p_{\ww};\mu_{\ww j},\rho_{\rm W}^{-1}\sigma^2_{\ww j})g(p_{\wm};\mu_{\wm j},\rho_{\rm M}^{-1}\sigma^2_{\wm j})
    \nonumber\\
    &=\int_0^1\int_0^1\{h(p'_{\ww})g(p_{\ww};f_{\rm W}^{\rm C}(p'_{\ww}),\rho_{\rm W}^{-1}s^2)g(p_{\wm};f_{\rm M}^{\rm C}(p'_{\wm}),\rho_{\rm M}^{-1}s^2)
    \nonumber\\
    &\hspace{1cm}+(1-h(p'_{\ww}))g(p_{\ww};f_{\rm W}^{\rm D}(p'_{\ww}),\rho_{\rm W}^{-1}s^2)g(p_{\wm};f_{\rm M}^{\rm D}(p'_{\wm}),\rho_{\rm M}^{-1}s^2)\}
    \nonumber\\
    &\hspace{1cm}\times\sum_jq_{{\rm W}j}g(p'_{\ww};\mu_{\ww j},\rho_{\rm W}^{-1}\sigma^2_{\ww j})g(p'_{\wm};\mu_{\wm j},\rho_{\rm M}^{-1}\sigma^2_{\wm j})\mathrm{d}p'_{\ww}\mathrm{d}p'_{\wm},
    \nonumber\\
    &=\sum_jq_{{\rm W}j}\int_0^1\int_0^1\{h(p'_{\ww})g(p_{\ww};f_{\rm W}^{\rm C}(p'_{\ww}),\rho_{\rm W}^{-1}s^2)g(p_{\wm};f_{\rm M}^{\rm C}(p'_{\wm}),\rho_{\rm M}^{-1}s^2)
    \nonumber\\
    &\hspace{1cm}+(1-h(p'_{\ww}))g(p_{\ww};f_{\rm W}^{\rm D}(p'_{\ww}),\rho_{\rm W}^{-1}s^2)g(p_{\wm};f_{\rm M}^{\rm D}(p'_{\wm}),\rho_{\rm M}^{-1}s^2)\}
    \nonumber\\
    &\hspace{1cm}\times g(p'_{\ww};\mu_{\ww j},\rho_{\rm W}^{-1}\sigma^2_{\ww j})g(p'_{\wm};\mu_{\wm j},\rho_{\rm M}^{-1}\sigma^2_{\wm j})\mathrm{d}p'_{\ww}\mathrm{d}p'_{\wm},
    \nonumber\\
    &\simeq\sum_jq_{{\rm W}j}\int_{-\infty}^{\infty}\int_{-\infty}^{\infty}\{h(\mu_{\ww j})g(p_{\ww};f_{\rm W}^{\rm C}(p'_{\ww}),\rho_{\rm W}^{-1}s^2)g(p_{\wm};f_{\rm M}^{\rm C}(p'_{\wm}),\rho_{\rm M}^{-1}s^2)
    \nonumber\\
    &\hspace{1cm}+(1-h(\mu_{\ww j}))g(p_{\ww};f_{\rm W}^{\rm D}(p'_{\ww}),\rho_{\rm W}^{-1}s^2)g(p_{\wm};f_{\rm M}^{\rm D}(p'_{\wm}),\rho_{\rm M}^{-1}s^2)\}
    \nonumber\\
    &\hspace{1cm}\times g(p'_{\ww};\mu_{\ww j},\rho_{\rm W}^{-1}\sigma^2_{\ww j})g(p'_{\wm};\mu_{\wm j},\rho_{\rm M}^{-1}\sigma^2_{\wm j})\mathrm{d}p'_{\ww}\mathrm{d}p'_{\wm},
    \nonumber\\
    &=\sum_jq_{{\rm W}j}\{h(\mu_{\ww j})g(p_{\ww};f_{\rm W}^{\rm C}(\mu_{\ww j}),\rho_{\rm W}^{-1}(s^2+(\Delta f_{\rm W}^{\rm C})^2\sigma_{\ww j}^2))g(p_{\wm};f_{\rm M}^{\rm C}(\mu_{\wm j}),\rho_{\rm M}^{-1}(s^2+(\Delta f_{\rm M}^{\rm C})^2\sigma_{\wm j}^2))\}
    \nonumber\\
    &\hspace{1cm}+(1-h(\mu_{\ww j}))g(p_{\ww};f_{\rm W}^{\rm D}(\mu_{\ww j}),\rho_{\rm W}^{-1}(s^2+(\Delta f_{\rm W}^{\rm D})^2\sigma_{\ww j}^2))g(p_{\wm};f_{\rm M}^{\rm D}(\mu_{\wm j}),\rho_{\rm M}^{-1}(s^2+(\Delta f_{\rm M}^{\rm D})^2\sigma_{\wm j}^2))\}.
    \label{equilibrium_W}
\end{align}
This equation gives a constraint for $(q_{{\rm W} j},\mu_{\ww j},\sigma_{\ww j}^2,\mu_{\wm j},\sigma_{\wm j}^2)$. Next, when $A={\rm M}$, in a similar manner, we obtain
\begin{align}
    &\sum_jq_{{\rm M}j}g(p_{\mw};\mu_{\mw j},\rho_{\rm W}^{-1}\sigma^2_{\mw j})g(p_{\mm};\mu_{\mm j},\rho_{\rm M}^{-1}\sigma^2_{\mm j})
    \nonumber\\
    &=\int_0^1\int_0^1\{h(p'_{\wm})g(p_{\mw};f_{\rm W}^{\rm C}(p'_{\mw}),\rho_{\rm W}^{-1}s^2)g(p_{\mm};f_{\rm M}^{\rm C}(p'_{\mm}),\rho_{\rm M}^{-1}s^2)
    \nonumber\\
    &\hspace{1cm}+(1-h(p'_{\wm}))g(p_{\mw};f_{\rm W}^{\rm D}(p'_{\mw}),\rho_{\rm W}^{-1}s^2)g(p_{\mm};f_{\rm M}^{\rm D}(p'_{\mm}),\rho_{\rm M}^{-1}s^2)\}
    \nonumber\\
    &\hspace{1cm}\times\sum_jq_{{\rm M}j}g(p'_{\mw};\mu_{\mw j},\rho_{\rm W}^{-1}\sigma^2_{\mw j})g(p'_{\mm};\mu_{\mm j},\rho_{\rm M}^{-1}\sigma^2_{\mm j})\mathrm{d}p'_{\mw}\mathrm{d}p'_{\mm}
    \nonumber\\
    &=\sum_jq_{{\rm M}j}\int_0^1\int_0^1\{h(p'_{\wm})g(p_{\mw};f_{\rm W}^{\rm C}(p'_{\mw}),\rho_{\rm W}^{-1}s^2)g(p_{\mm};f_{\rm M}^{\rm C}(p'_{\mm}),\rho_{\rm M}^{-1}s^2)
    \nonumber\\
    &\hspace{1cm}+(1-h(p'_{\wm}))g(p_{\mw};f_{\rm W}^{\rm D}(p'_{\mw}),\rho_{\rm W}^{-1}s^2)g(p_{\mm};f_{\rm M}^{\rm D}(p'_{\mm}),\rho_{\rm M}^{-1}s^2)\}
    \nonumber\\
    &\hspace{1cm}\times\sum_jq_{{\rm M}j}g(p'_{\mw};\mu_{\mw j},\rho_{\rm W}^{-1}\sigma^2_{\mw j})g(p'_{\mm};\mu_{\mm j},\rho_{\rm M}^{-1}\sigma^2_{\mm j})\mathrm{d}p'_{\mw}\mathrm{d}p'_{\mm}
    \nonumber\\
    &\simeq\sum_jq_{{\rm M}j}\int_{-\infty}^{\infty}\int_{-\infty}^{\infty}\{h(\mu_{\wm j})g(p_{\mw};f_{\rm W}^{\rm C}(p'_{\mw}),\rho_{\rm W}^{-1}s^2)g(p_{\mm};f_{\rm M}^{\rm C}(p'_{\mm}),\rho_{\rm M}^{-1}s^2)
    \nonumber\\
    &\hspace{1cm}+(1-h(\mu_{\wm j}))g(p_{\mw};f_{\rm W}^{\rm D}(p'_{\mw}),\rho_{\rm W}^{-1}s^2)g(p_{\mm};f_{\rm M}^{\rm D}(p'_{\mm}),\rho_{\rm M}^{-1}s^2)\}
    \nonumber\\
    &\hspace{1cm}\times\sum_jq_{{\rm M}j}g(p'_{\mw};\mu_{\mw j},\rho_{\rm W}^{-1}\sigma^2_{\mw j})g(p'_{\mm};\mu_{\mm j},\rho_{\rm M}^{-1}\sigma^2_{\mm j})\mathrm{d}p'_{\mw}\mathrm{d}p'_{\mm}
    \nonumber\\
    &=\sum_jq_{{\rm M}j}\{h(\mu_{\wm j})g(p_{\mw};f_{\rm W}^{\rm C}(\mu_{\mw j}),\rho_{\rm W}^{-1}(s^2+(\Delta f_{\rm W}^{\rm C})^2\sigma_{\mw j}^2))g(p_{\mm};f_{\rm M}^{\rm C}(\mu_{\mm j}),\rho_{\rm M}^{-1}(s^2+(\Delta f_{\rm M}^{\rm C})^2\sigma_{\mm j}^2))\}
    \nonumber\\
    &\hspace{1cm}+(1-h(\mu_{\wm j}))g(p_{\mw};f_{\rm W}^{\rm D}(\mu_{\mw j}),\rho_{\rm W}^{-1}(s^2+(\Delta f_{\rm W}^{\rm D})^2\sigma_{\mw j}^2))g(p_{\mm};f_{\rm M}^{\rm D}(\mu_{\mm j}),\rho_{\rm M}^{-1}(s^2+(\Delta f_{\rm M}^{\rm D})^2\sigma_{\mm j}^2))\}.
    \label{equilibrium_M}
\end{align}
This equation gives a constraint for $(q_{{\rm M} j},\mu_{\mw j},\sigma_{\mw j}^2,\mu_{\mm j},\sigma_{\mm j}^2)$.

To solve \eqref{equilibrium_W}, let us consider a set of solutions $\{(\mu_{\ww j},\mu_{\wm j})\}_{j}$. From the equilibrium condition of \eqref{equilibrium_W}, the equal set must be restored by applying C-map and D-map to all the elements of the set. In other words, the condition is given by
\begin{align}
    \{(\mu_{\ww j},\mu_{\wm j})\}_{j}=\{(f_{\rm W}^{\rm C}(\mu_{\ww j}),f_{\rm M}^{\rm C}(\mu_{\wm j}))\}_{j}\cup\{(f_{\rm W}^{\rm D}(\mu_{\ww j}),f_{\rm M}^{\rm D}(\mu_{\wm j}))\}_{j}.
    \label{set_problem_W}
\end{align}
Similarly, to obtain the equilibrium condition for \eqref{equilibrium_M}, we should consider a set of solutions $\{(\mu_{\mw j},\mu_{\mm j})\}_{j}$ satisfying
\begin{align}
    \{(\mu_{\mw j},\mu_{\mm j})\}_{j}=\{(f_{\rm W}^{\rm C}(\mu_{\mw j}),f_{\rm M}^{\rm C}(\mu_{\mm j}))\}_{j}\cup\{(f_{\rm W}^{\rm D}(\mu_{\mw j}),f_{\rm M}^{\rm D}(\mu_{\mm j}))\}_{j}.
    \label{set_problem_M}
\end{align}
Here, although the appearance of the variables are different, the problems are essentially between \eqref{set_problem_W} and \eqref{set_problem_M}. Thus, the problem to be solved is
\begin{align}
    \{(\mu_{j, {\rm W}},\mu_{j, {\rm M}})\}_{j}=\{(f_{\rm W}^{\rm C}(\mu_{j, {\rm W}}),f_{\rm M}^{\rm C}(\mu_{j, {\rm M}}))\}_{j}\cup\{(f_{\rm W}^{\rm D}(\mu_{j, {\rm W}}),f_{\rm M}^{\rm D}(\mu_{j, {\rm M}}))\}_{j}.
    \label{set_problem}
\end{align}
Furthermore, because ${\rm W},{\rm M}\in{\cal S}$, we can generalize the problem as
\begin{align}
    \{(\mu_{j, S_{01}},\cdots,\mu_{j, S_{16}})\}_{j}=\{(f_{S_{01}}^{\rm C}(\mu_{j, S_{01}}),\cdots,f_{S_{16}}^{\rm C}(\mu_{j, S_{16}}))\}_{j}\cup\{(f_{S_{01}}^{\rm D}(\mu_{j, S_{01}}),\cdots,f_{S_{16}}^{\rm D}(\mu_{j, S_{16}}))\}_{j}.
    \label{set_problem_gen}
\end{align}

Now, for all $S_k\in{\cal S}$ and, let us consider a set $\{\mu_{j,S_k}\}_{j\in\mathbb{Z}\backslash\{0\}}$ satisfying
\begin{align}
    \mu_{+1,S_k}&=f_{S_k}^{\rm C}(\mu_{-1,S_k})=f_{S_k}^{\rm C}(\mu_{-2,S_k})=\cdots,
    \label{numbering_a}\\
    \mu_{+(j+1),S_k}&=f_{S_k}^{\rm C}(\mu_{+j,S_k}),
    \label{numbering_b}\\
    \mu_{-1,S_k}&=f_{S_k}^{\rm D}(\mu_{+1,S_k})=f_{S_k}^{\rm D}(\mu_{+2,S_k})=\cdots,
    \label{numbering_c}\\
    \mu_{-(j+1),S_k}&=f_{S_k}^{\rm D}(\mu_{-j,S_k}),
    \label{numbering_d}
\end{align}
(the proof for these equations will be given later). This set $\{\mu_{j,S_k}\}_{j\in\mathbb{Z}\backslash\{0\}}$ gives a solution to problem \eqref{set_problem_gen}, and thus solves \eqref{equilibrium_W} and \eqref{equilibrium_M}. In \eqref{numbering_a}-\eqref{numbering_d}, we consistently label each $\mu_{j,S_k}$ such that sequentially applying C-map (resp. D-map) $j(>0)$ times leads to label $+j$ (resp. $-j$) (see the illustration in Fig.~\ref{FS01}). In the following, we show that such $\{\mu_{j,S_k}\}_{j\in\mathbb{Z}\backslash\{0\}}$ actually exists for all $k$.

{\bf 1. When neither C-map nor D-map is constant:} First, we consider a case of $\Delta f_{S_k}^{\rm C}\neq 0$ and $\Delta f_{S_k}^{\rm D}\neq 0$. Four norms of $k=06,07,09,10$ correspond to this case. In this case, C-map and D-map have the same fixed point (at $1/2$). Only the position given by this fixed point is achieved at an equilibrium. Indeed, if we substitute $\mu_{j,S_k}=1/2$ for all $j\in \mathbb{Z}\backslash\{0\}$, \eqref{numbering_a}-\eqref{numbering_d} are simultaneously satisfied without any contradiction.

{\bf 2. When both C-map and D-map are constant:} Second, we consider a case of $\Delta f_{S_k}^{\rm C}=0$ and $\Delta f_{S_k}^{\rm D}=0$. Four norms of $k=01,04,13,16$ correspond to this case. Because $f_{S_k}^{\rm C}$ is a constant map, \eqref{numbering_a} and \eqref{numbering_b} are satisfied by substituting the mapped value of this map into $\mu_{+j,S_k}$ for all $j=1,2,\cdots$. In the same way, because $f_{S_k}^{\rm D}$ is a constant map, \eqref{numbering_c} and \eqref{numbering_d} are satisfied by substituting the mapped value into $\mu_{-j,S_k}$ for all $j=1,2,\cdots$. Thus, no contradiction occurs.

{\bf 3. When only C-map is constant:} Third, we consider a case of $\Delta f_{S_k}^{\rm C}=0$ and $\Delta f_{S_k}^{\rm D}\neq 0$. Four norms $k=02,03,14,15$ correspond to this case. Because $f_{S_k}^{\rm C}$ is a constant map, \eqref{numbering_a} and \eqref{numbering_b} are satisfied by substituting the mapped value of this map into $\mu_{+j,S_k}$ for all $j=1,2,\cdots$. Then, we define $\mu_{-1,S_k}$ as the value to which D-map maps all the same value $\mu_{+1,S_k}=\mu_{+2,S_k}=\cdots$, and \eqref{numbering_c} is satisfied. Finally, we sequentially define $\mu_{-2,S_k},\mu_{-3,S_k},\cdots$ by applying D-map to $\mu_{-1,S_k}$ one by one. Thus, no contradiction occurs.

{\bf 4. When only D-map is constant:} Finally, we consider a case of $\Delta f_{S_k}^{\rm C}\neq 0$ and $\Delta f_{S_k}^{\rm D}=0$. Four norms $k=05,08,09,12$ correspond to this case. Because $f_{S_k}^{\rm D}$ is a constant map, \eqref{numbering_c}, \eqref{numbering_d} are satisfied by substituting the mapped value of this map into $\mu_{-j,S_k}$ for all $j=1,2,\cdots$. Then, we define $\mu_{+1,S_k}$ as the value to which D-map maps all the same value $\mu_{-1,S_k}=\mu_{-2,S_k}=\cdots$, and \eqref{numbering_c} is satisfied. Finally, we sequentially define $\mu_{+2,S_k},\mu_{+3,S_k},\cdots$ by applying C-map to $\mu_{+1,S_k}$ one by one. Thus, no contradiction occurs.

As summarized in Table~\ref{T01}, the set $\{\mu_{j,S_k}\}_{j\in\mathbb{Z}\backslash\{0\}}$ can be analytically described. Furthermore, we also define $\sigma_{j,S_k}^2$ as the variance in Gaussian corresponding to the mean $\mu_{j,S_k}$. Similarly to the mean values above, we solve the variances as
\begin{align}
    (\sigma_{\ww j}^2,\sigma_{\wm j}^2)=(\sigma_{\mw j}^2,\sigma_{\mm j}^2)=(\sigma_{j,{\rm W}}^2,\sigma_{j,{\rm M}}^2).
\end{align}
The recursion that the set $\{\sigma_{j,S_k}^2\}_{j\in\mathbb{Z}\backslash\{0\}}$ should satisfy is
\begin{align}
    \sigma_{+1,S_k}^2&=s^2+(\Delta f_{S_k}^{\rm C})^2\sigma_{-1,S_k}^2=s^2+(\Delta f_{S_k}^{\rm C})^2\sigma_{-2,S_k}^2=\cdots,\\
    \sigma_{+(j+1),S_k}^2&=s^2+(\Delta f_{S_k}^{\rm C})^2\sigma_{+j,S_k}^2,\\
    \sigma_{-1,S_k}^2&=s^2+(\Delta f_{S_k}^{\rm D})^2\sigma_{+1,S_k}^2=s^2+(\Delta f_{S_k}^{\rm D})^2\sigma_{+2,S_k}^2=\cdots,\\
    \sigma_{-(j+1),S_k}^2&=s^2+(\Delta f_{S_k}^{\rm D})^2\sigma_{-j,S_k}^2,
\end{align}
and the solution exists for all $S_k$ (see Table.~\ref{T01} for the solution of these equations). Table.~\ref{T01} shows $\{(\mu_{j,S_k},\sigma_{j,S_k}^2)\}_{j\in\mathbb{Z}\backslash\{0\}}$.
%\begin{landscape}
% Table 01
\renewcommand{\arraystretch}{2.0}
\begin{table}[htbp]
\centering
\begin{tabular}{c|c|c|c|c|}
    $S_k$ & $\mu_{+j,S_k}$ & $\mu_{-j,S_k}$ & $\sigma_{+j,S_k}^2$ & $\sigma_{-j,S_k}^2$ \\
    \hline
    $S_{01}$ & $1-e_2$ & $1-e_2$ & $\displaystyle\frac{e_2(1-e_2)}{N}$ & $\displaystyle\frac{e_2(1-e_2)}{N}$ \\
    $S_{02}$ & $1-e_2$ & $\displaystyle\frac{1+(1-2e_2)^{j+1}}{2}$ & $\displaystyle\frac{e_2(1-e_2)}{N}$ & $\displaystyle\frac{1-(1-2e_2)^{2(j+1)}}{4N}$ \\
    $S_{03}$ & $1-e_2$ & $\displaystyle\frac{1-\{-(1-2e_2)\}^{j+1}}{2}$ & $\displaystyle\frac{e_2(1-e_2)}{N}$ & $\displaystyle\frac{1-(1-2e_2)^{2(j+1)}}{4N}$ \\
    $S_{04}$ & $1-e_2$ & $e_2$ & $\displaystyle\frac{e_2(1-e_2)}{N}$ & $\displaystyle\frac{e_2(1-e_2)}{N}$ \\
    $S_{05}$ & $\displaystyle\frac{1+(1-2e_2)^{j+1}}{2}$ & $1-e_2$ & $\displaystyle\frac{1-(1-2e_2)^{2(j+1)}}{4N}$ & $\displaystyle\frac{e_2(1-e_2)}{N}$ \\
    $S_{07}$ & $\displaystyle\frac{1}{2}$ & $\displaystyle\frac{1}{2}$ & $\displaystyle\frac{1}{4N}$ & $\displaystyle\frac{1}{4N}$ \\
    $S_{08}$ & $\displaystyle\frac{1-(1-2e_2)^{j+1}}{2}$ & $e_2$ & $\displaystyle\frac{1-(1-2e_2)^{2(j+1)}}{4N}$ & $\displaystyle\frac{e_2(1-e_2)}{N}$ \\
    $S_{09}$ & $\displaystyle\frac{1-\{-(1-2e_2)\}^{j+1}}{2}$ & $1-e_2$ & $\displaystyle\frac{1-(1-2e_2)^{2(j+1)}}{4N}$ & $\displaystyle\frac{e_2(1-e_2)}{N}$ \\
    $S_{12}$ & $\displaystyle\frac{1+\{-(1-2e_2)\}^{j+1}}{2}$ & $e_2$ & $\displaystyle\frac{1-(1-2e_2)^{2(j+1)}}{4N}$ & $\displaystyle\frac{e_2(1-e_2)}{N}$ \\
    $S_{13}$ & $e_2$ & $1-e_2$ & $\displaystyle\frac{e_2(1-e_2)}{N}$ & $\displaystyle\frac{e_2(1-e_2)}{N}$ \\
    $S_{14}$ & $e_2$ & $\displaystyle\frac{1-(1-2e_2)^{j+1}}{2}$ & $\displaystyle\frac{e_2(1-e_2)}{N}$ & $\displaystyle\frac{1-(1-2e_2)^{2(j+1)}}{4N}$ \\
    $S_{15}$ & $e_2$ & $\displaystyle\frac{1+\{-(1-2e_2)\}^{j+1}}{2}$ & $\displaystyle\frac{e_2(1-e_2)}{N}$ & $\displaystyle\frac{1-(1-2e_2)^{2(j+1)}}{4N}$ \\
    $S_{16}$ & $e_2$ & $e_2$ & $\displaystyle\frac{e_2(1-e_2)}{N}$ & $\displaystyle\frac{e_2(1-e_2)}{N}$ \\
    \hline
\end{tabular}
\caption{Analytical solution of Gaussian functions. We omit $S_{06}$, $S_{10}$, and $S_{11}$ because the results are identical to those of $S_{07}$(SJ).}
\label{T01}
\end{table}
\renewcommand{\arraystretch}{1.0}
%\end{landscape}

We also calculate a set of the masses of Gaussian functions, i.e., $\{q_{{\rm W}j}\}_{j\in\mathbb{Z}\backslash\{0\}}$ and $\{q_{{\rm M}j}\}_{j\in\mathbb{Z}\backslash\{0\}}$. By substituting the values in Table~\ref{T01} into \eqref{equilibrium_W}, we obtain the following relational expressions
\begin{align}
    &q_{{\rm W}+1}=\sum_{j=1}^{\infty}h(\mu_{-j,{\rm W}})q_{{\rm W}-j},
    \label{mass_a}\\
    &q_{{\rm W}+j}=h(\mu_{+(j-1),{\rm W}})q_{{\rm W}+(j-1)}\quad (j=2,\cdots,\infty),
    \label{mass_b}\\
    &q_{{\rm W}-1}=\sum_{j=1}^{\infty}(1-h(\mu_{+j,{\rm W}}))q_{{\rm W}+j},
    \label{mass_c}\\
    &q_{{\rm W}-j}=(1-h(\mu_{-(j-1),{\rm W}}))q_{{\rm W}-(j-1)}\quad (j=2,\cdots,\infty).
    \label{mass_d}
\end{align}
Similarly, by substituting the values in Table~\ref{T01} into \eqref{equilibrium_M}, we obtain
\begin{align}
    &q_{{\rm M}+1}=\sum_{j=1}^{\infty}h(\mu_{-j,{\rm M}})q_{{\rm W}-j},
    \label{mass_e}\\
    &q_{{\rm M}+j}=h(\mu_{+(j-1),{\rm M}})q_{{\rm W}+(j-1)}\quad (j=2,\cdots,\infty),
    \label{mass_f}\\
    &q_{{\rm M}-1}=\sum_{j=1}^{\infty}(1-h(\mu_{+j,{\rm M}}))q_{{\rm W}+j},
    \label{mass_g}\\
    &q_{{\rm M}-j}=(1-h(\mu_{-(j-1),{\rm M}}))q_{{\rm W}-(j-1)}\quad (j=2,\cdots,\infty).
    \label{mass_h}
\end{align}
\eqref{mass_a}-\eqref{mass_h} includes the infinite summations. Because these infinite summation cannot be analytically calculated, one should set a cutoff of the summations in the numerical calculation of \eqref{mass_a}-\eqref{mass_h}.

Fig.~2-B in the main manuscript shows an example of $\Phi_{\rm W}^*(p_{\ww},p_{\wm})$. In this example, the elements in $\{(\mu_{j,{\rm W}},\mu_{j,{\rm M}})\}_{j\in\mathbb{Z}\backslash\{0\}}$ are all different for different $j$, and thus the labeling in this study is at least necessary for the description of the reputation structure. This figure also shows that the obtained analytical solutions well approximate simulations of the image matrix.

\section{Calculation of expected payoff}
In order to consider an evolutionary process, we derive expected payoffs of wild-types ${\rm W}$ and mutants ${\rm M}$ from joint probability distribution of goodnesses, i.e., $\Phi^{*}_{\rm W}$ and $\Phi^{*}_{\rm M}$. In the limit that mutants are rare $\delta\to 0$, the expected payoffs of the wild-types $u_{\rm W}$ and mutants $u_{\rm M}$ are given by
\begin{equation}
\begin{split}
    u_{\rm W}&=(b-c)\bar{p}_{\ww},\\
    u_{\rm M}&=b\bar{p}_{\mw}-c\bar{p}_{\wm},\\
\end{split}
\end{equation}
Here, $\bar{p}_{AA'}$ indicates the average goodnesses of $p_{AA'}$, i.e., described as
\begin{equation}
\begin{split}
    \bar{p}_{\ww}&=\int_{0}^{1}\int_{0}^{1}p_{\ww}\Phi^{*}_{\rm W}(p_{\ww},p_{\wm}){\rm d}p_{\ww}{\rm d}p_{\wm},\\
    \bar{p}_{\wm}&=\int_{0}^{1}\int_{0}^{1}p_{\wm}\Phi^{*}_{\rm W}(p_{\ww},p_{\wm}){\rm d}p_{\ww}{\rm d}p_{\wm},\\
    \bar{p}_{\mw}&=\int_{0}^{1}\int_{0}^{1}p_{\mw}\Phi^{*}_{\rm M}(p_{\mw},p_{\mm}){\rm d}p_{\mw}{\rm d}p_{\mm},\\
\end{split}
\end{equation}
This average goodness can be analytically calculated by Gaussian approximation of $\Phi^{*}_{A}(p_{A{\rm W}},p_{A{\rm M}})$. According to the conditions for the ESS, mutants can invade the population of wild-types if $u_{\rm W}>u_{\rm M}$.

\begin{figure}[htbp]
    \centering
    \includegraphics[width=0.85\hsize]{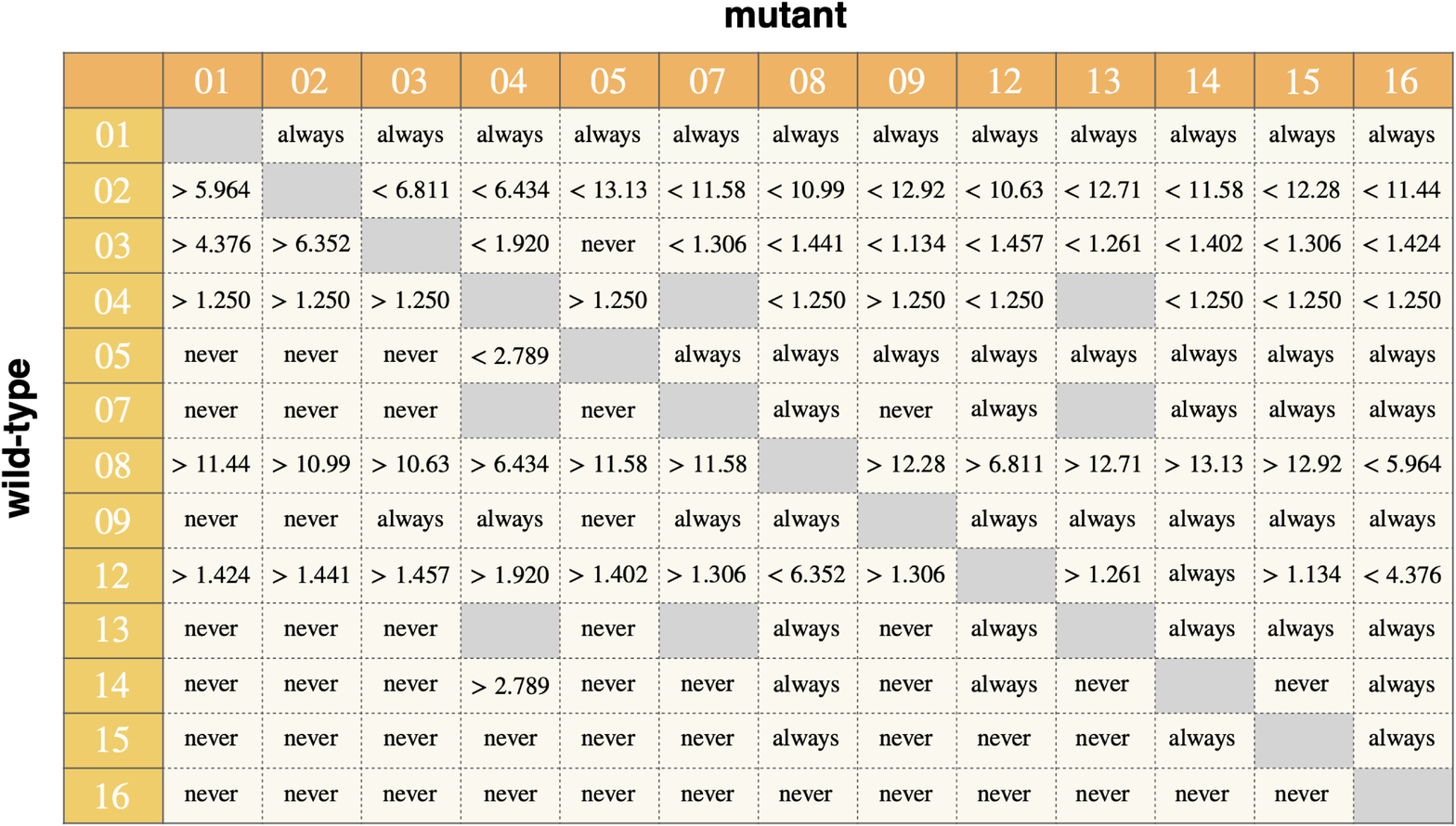}
    \caption{Regions for possible invasions in the evolutionary processes. The row and column indicate the wild-type and mutant norms, respectively. The matrix shows the region of $b/c(>1)$ where the mutant invades the wild-type. In some pairs of wild-type and mutant norms, the mutant always or never succeeds in invading the wild-types for all $b/c(>1)$. The calculation is based on $e_1=0$ and $e_2=0.1$.}
    \label{FS02}
\end{figure}

Regions where a mutant norm can invade a wild-type norm are given by Fig.~\ref{FS02}. From this figure, we can obtain the invasibilities for a certain $b/c$, as shown in Fig.~3-A in the main manuscript.
% Figure S02

\clearpage

\section{Calculation of equilibrium state in public reputation}
In this section, we derive the equilibrium distribution of reputations under public assessment, based on the previous study~\cite{uchida2010competition}. The basic setting is the same whether the reputation is publicly shared or privately held. We assume a population of size $N$ which consists of mutants with norm ${\rm M}$ and wild-type individuals with norm ${\rm W}\neq {\rm M}$. A donor and a recipient are randomly chosen every round. The donor chooses cooperation to the good recipient and defection to the bad recipient. Here, the donor erroneously chooses the opposite action to the intended one with probability $0\le e_1< 1/2$. Then, all the individuals update their reputations of the donor. The difference between the public and private reputation cases is seen in the observers' ways to update reputations. We assume that one mutant observer and one wild-type observer are chosen as representatives of each norm, and each gives a good or bad reputation to the donor according to its norm. Here, each representative observer commits an assignment error independently, in which case it erroneously assigns the opposite reputation to the intended one with probability $0<e_2<1/2$. (such an assessment error was not assumed in~\cite{uchida2010competition}) Then, all the individuals with the same norm copy the reputation of the donor assigned by their representative. Thus, the reputation of the same individual, even an erroneously assigned one, is shared among all the individuals with the same norm. In other words, each individual at any given time has two reputations, one is shared by all the mutant individuals, and the other is shared by all the wild-type individuals in the population.

Here we specifically consider the situation where rare mutants with norm ${\rm M}=S_{16}$(ALLB) invades a wild-type population with norm ${\rm W}\neq {\rm M}$. We use the same definition of $p_{AA'}$, i.e., goodness of an individual with norm $A$ in the eyes of norm $A'$ users. Because reputations are public, $p_{AA'}$ can be either $1$ (the individual is assigned as good from all) or $0$ (the individual is assigned as bad from all). Below we will derive $\bar{p}_{AA'}$, the probability that a norm $A$ user has a good reputation in the eyes of norm $A'$ users. 

Since the mutant norm is ALLB, the probability that mutants assign a good reputation to the donor is always $a_{\rm M}^{\rm GC}=a_{\rm M}^{\rm BC}=a_{\rm M}^{\rm GD}=a_{\rm M}^{\rm BD}= e_2$. Thus we obtain $\bar{p}_{\wm}=\bar{p}_{\mm}=e_2$. 

Next we aim to solve the equilibrium average goodnesses in the eyes of wild-types, $\bar{p}_{\ww}$ and $\bar{p}_{\mw}$. First, let us calculate $\bar{p}_{\ww}$, which is relevant when the donor and the observer use norm $\rm W$. Note that we can assume that the recipient uses norm $\rm W$, because mutants are rare. $\bar{p}_{\ww}$ should satisfy
\begin{equation}
    \bar{p}_{\ww}=\bar{p}_{\ww}\{(1-e_1)a_{\rm W}^{\rm GC}+e_1a_{\rm W}^{\rm GD}\}+(1-\bar{p}_{\ww})\{e_1a_{\rm W}^{\rm BC}+(1-e_1)a_{\rm W}^{\rm BD}\},
\end{equation}
The equality between the left- and right-hand sides shows that the proportion of good individuals balances before and after updating the chosen donor's reputation. In the right-hand side, $\bar{p}_{\ww}$ and $1-\bar{p}_{\ww}$ in the first and the second terms indicate the probabilities that a randomly chosen recipient of norm $\rm W$ is good or bad from the viewpoint of norm $\rm W$, respectively. When the recipient is good, the donor chooses cooperation or defection with probabilities $(1-e_1)$ and $e_1$. Then, $a_{\rm W}^{\rm GC}$ and $a_{\rm W}^{\rm GD}$ indicate the probabilities that the cooperating or defecting donor receives a good reputation from observers of norm $\rm W$. When the recipient is bad, the donor chooses cooperation or defection with probabilities $e_1$ and $(1-e_1)$. Then, $a_{\rm W}^{\rm BC}$ and $a_{\rm W}^{\rm BD}$ indicate the probabilities that the cooperating or defecting donor receives a good reputation from observers of norm $\rm W$. The solution is
\begin{equation}
    \bar{p}_{\ww}=\frac{(1-e_1)a_{\rm W}^{\rm BD}+e_1a_{\rm W}^{\rm BC}}{1-\{(1-e_1)(a_{\rm W}^{\rm GC}-a_{\rm W}^{\rm BD})+e_1(a_{\rm W}^{\rm GD}-a_{\rm W}^{\rm BC})\}}.
\end{equation}

Second, let us calculate $\bar{p}_{\mw}$, which is relevant when the donor uses norm ${\rm M}$ and the observer uses norm ${\rm W}$. Note that we can once again assume that the recipient uses norm $\rm W$ because mutants are rare. $\bar{p}_{\mw}$ should satisfy
\begin{align}
    &\bar{p}_{\mw}=\bar{p}_{\ww}\{h(e_2)a_{\rm W}^{\rm GC}+(1-h(e_2))a_{\rm W}^{\rm GD}\}+(1-\bar{p}_{\ww})\{h(e_2)a_{\rm W}^{\rm BC}+(1-h(e_2))a_{\rm W}^{\rm BD}\}.
\end{align}
Here, $\bar{p}_{\ww}$ and $(1-\bar{p}_{\ww})$ in the first and second terms of the right-hand side indicate the probabilities that the recipient is good or bad from the viewpoint of norm ${\rm W}$, respectively. In both terms, $h(e_2) (=e_2 (1-e_1)+(1-e_2)e_1)$ and $1-h(e_2)$ are the probabilities that the donor with norm $\rm M$ executes cooperation or defection, which is independent of whether the recipient is good or bad from the viewpoint of norm $\rm W$. In the first term, $a_{\rm W}^{\rm GC}$ and $a_{\rm W}^{\rm GD}$ indicate the probabilities that the cooperating and defecting donor receives a good reputation from the observers of norm ${\rm W}$. In the second term, $a_{\rm W}^{\rm BC}$ and $a_{\rm W}^{\rm BD}$ indicate the probabilities that the cooperating and defecting donor receives a good reputation from the observers of norm ${\rm W}$.

We summarize the solutions, $\bar{p}_{\ww}$ and $\bar{p}_{\mw}$, in Table~\ref{T02}.
% Table 02
\begin{landscape}
\renewcommand{\arraystretch}{2.0}
\begin{table}[htbp]
\centering
\begin{tabular}{c|c|c|c||c|c|}
    $S_k$ & $\bar{p}_{\ww}$ & & $\bar{p}_{\mw}$ & $\bar{p}_{\ww}|_{e_1=0}$ & $\bar{p}_{\mw}|_{e_1=0}$ \\
    \hline
    $S_{01}$ & $1-e_2$ & $=$ & $1-e_2$ & $1-e_2$ & $1-e_2$ \\
    $S_{02}$ & $\displaystyle\frac{e_1+e_2-2e_1e_2}{e_1+2e_2-2e_1e_2}$ & $<$ & $\displaystyle\frac{e_1+e_2-2e_1e_2+e_2^2-2e_1e_2^2-2e_2^3+4e_1e_2^3}{e_1+2e_2-2e_1e_2}$ & $\displaystyle\frac{1}{2}$ & $\displaystyle\frac{1}{2}+\frac{1}{2}e_2-e_2^2$ \\
    $S_{03}$ & $\displaystyle\frac{1-e_2}{1+e_1-2e_1e_2}$ & $>$ & $\displaystyle\frac{(1-e_2)(2e_1+3e_2-6e_1e_2-2e_2^2+4e_1e_2^2)}{1+e_1-2e_1e_2}$ & $1-e_2$ & $3e_2-5e_2^2+2e_2^3$ \\
    $S_{04}$ & $\displaystyle\frac{1}{2}$ & $>$ & $\displaystyle e_1+2e_2-4e_1e_2-2e_2^2+4e_1e_2^2$ & $\displaystyle\frac{1}{2}$ & $2e_2-2e_2^2$ \\
    $S_{05}$ & $\displaystyle\frac{1-e_1-e_2+2e_1e_2}{1-e_1+2e_1e_2}$ & $>$ & $\displaystyle\frac{1-e_1-e_2+2e_1e_2-e_2^2+2e_1e_2^2+2e_2^3-4e_1e_2^3}{1-e_1+2e_1e_2}$ & $1-e_2$ & $1-e_2-e_2^2+2e_2^3$ \\
    $S_{06}$ & $\displaystyle\frac{1}{2}$ & $=$ & $\displaystyle\frac{1}{2}$ & $\displaystyle\frac{1}{2}$ & $\displaystyle\frac{1}{2}$ \\
    $S_{07}$ & $\displaystyle 1-e_1-e_2+2e_1e_2$ & $>$ & $\displaystyle 2e_1+3e_2-2e_1^2-12e_1e_2-6e_2^2+12e_1^2e_2+24e_1e_2^2+4e_2^3-24e_1^2e_2^2-16e_1e_2^3+16e_1^2e_2^3$ & $1-e_2$ & $3e_2-6e_2^2+4e_2^3$ \\
    $S_{08}$ & $\displaystyle\frac{e_2}{e_1+2e_2-2e_1e_2}$ & $>$ & $\displaystyle\frac{e_2(2e_1+3e_2-6e_1e_2-2e_2^2+4e_1e_2^2)}{e_1+2e_2-2e_1e_2}$ & $\displaystyle\frac{1}{2}$ & $\displaystyle\frac{3}{2}e_2-e_2^2$ \\
    $S_{09}$ & $\displaystyle\frac{1-e_2}{2-e_1-2e_2+2e_1e_2}$ & $<$ & $\displaystyle\frac{(1-e_2)(2-2e_1-3e_2+6e_1e_2+2e_2^2-4e_1e_2^2)}{2-e_1-2e_2+2e_1e_2}$ & $\displaystyle\frac{1}{2}$ & $\displaystyle 1-\frac{3}{2}e_2+e_2^2$ \\
    $S_{10}$ & $\displaystyle e_1+e_2-2e_1e_2$ & $<$ & $\displaystyle 2e_1+3e_2-2e_1^2-12e_1e_2-6e_2^2+12e_1^2e_2+24e_1e_2^2+4e_2^3-24e_1^2e_2^2-16e_1e_2^3+16e_1^2e_2^3$ & $e_2$ & $3e_2-6e_2^2+4e_2^3$ \\
    $S_{11}$ & $\displaystyle\frac{1}{2}$ & $=$ & $\displaystyle\frac{1}{2}$ & $\displaystyle\frac{1}{2}$ & $\displaystyle\frac{1}{2}$ \\
    $S_{12}$ & $\displaystyle\frac{e_1+e_2-2e_1e_2}{1+e_1-2e_1e_2}$ & $<$ & $\displaystyle\frac{e_1+2e_2-4e_1e_2-3e_2^2+6e_1e_2^2+2e_2^3-4e_1e_2^3}{1+e_1-2e_1e_2}$ & $e_2$ & $2e_2-3e_2^2+2e_2^3$ \\
    $S_{13}$ & $\displaystyle\frac{1}{2}$ & $<$ & $\displaystyle 1-e_1-2e_2+4e_1e_2+2e_2^2-4e_1e_2^2$ & $\displaystyle\frac{1}{2}$ & $1-2e_2+2e_2^2$ \\
    $S_{14}$ & $\displaystyle\frac{e_2}{1-e_1+2e_1e_2}$ & $<$ & $\displaystyle\frac{e_2(2-2e_1-3e_2+6e_1e_2+2e_2^2-4e_1e_2^2)}{1-e_1+2e_1e_2}$ & $e_2$ & $2e_2-3e_2^2+2e_2^3$ \\
    $S_{15}$ & $\displaystyle\frac{1-e_1-e_2+2e_1e_2}{2-e_1-2e_2+2e_1e_2}$ & $>$ & $\displaystyle\frac{1-e_1-2e_2+4e_1e_2+3e_2^2-6e_1e_2^2-2e_2^3+4e_1e_2^3}{2-e_1-2e_2+2e_1e_2}$ & $\displaystyle\frac{1}{2}$ & $\displaystyle\frac{1}{2}-\frac{1}{2}e_2+e_2^2$ \\
    $S_{16}$ & $\displaystyle e_2$ & $=$ & $\displaystyle e_2$ & $e_2$ & $e_2$ \\
    \hline
\end{tabular}
\caption{Analytical solution of reputation structure under public assessment. Here, the equality (i.e., $=$) and inequality (i.e., $>$ or $<$) signs show the relations between $\bar{p}_{\ww}$ and $\bar{p}_{\mw}$ for all of $0\le e_1<1/2$ and $0<e_2<1/2$.} 
\label{T02}
\end{table}
\renewcommand{\arraystretch}{1.0}
\end{landscape}

\begin{figure}[htbp]
    \centering
    \includegraphics[width=0.45\hsize]{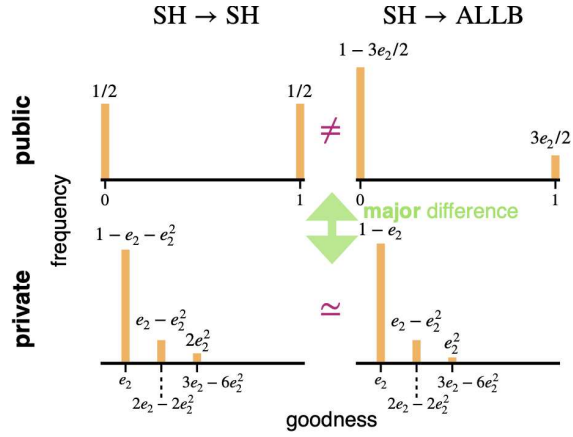}
    \caption{Illustration of how the wild-type SH gives reputations to the self and mutant ALLB norms. In each panel, the horizontal and vertical axes indicate the goodness and its frequency, respectively. Positions and heights of bars are correct only up to order $e_2^2$. By comparing the upper panels with the lower ones, we can see that the reputation from SH differs significantly between the public and private reputation cases. In the private reputation case, SH still manages to distinguish the self norm with ALLB, but only with the difference of order of $e_2^2$.}
    \label{FS03}
\end{figure}

Based on Table~\ref{T02}, we can see how the reputation structure differs between the public and private reputation cases. The reputation structure for norms $S_{03}$(SS) and $S_{07}$(SJ) are illustrated in Fig.~3-D and E in the main manuscript, while that of $S_{08}$(SH) is in Fig.~\ref{FS03}.
% Figure S03

\clearpage

\section{Numerical algorithm and error estimate}
This section provides how to computationally calculate \eqref{mass_a}-\eqref{mass_d} and \eqref{mass_e}-\eqref{mass_h} with sufficient accuracy. 

\begin{figure}[htbp]
    \centering
    \includegraphics[width=0.8\hsize]{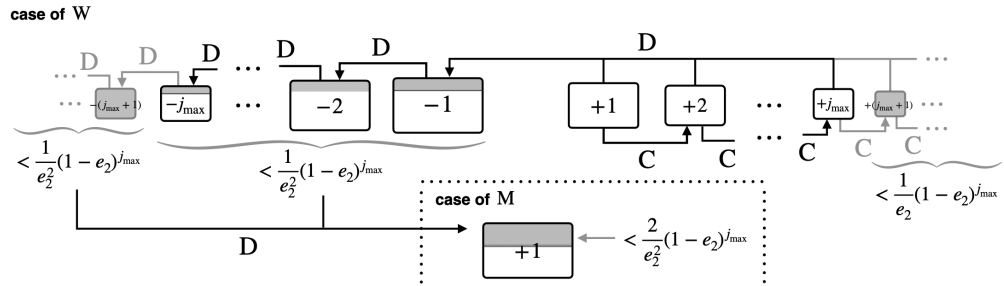}
    \caption{An illustration of numerical algorithm and error estimation of the masses. The black and gray arrows show how theoretical calculations of \eqref{mass_a'}-\eqref{mass_d'} are performed, whereas only black arrows are relevant in the computation of \eqref{mass_a''}-\eqref{mass_d2''}. Each box shows the size of $Q_{{\rm W}j}$. The gray part in each box shows the size of approximation error, $Q_{{\rm W}j}-\hat{Q}_{{\rm W}j}$. Apart from $Q_{{\rm W}j}$, the area surrounded by dots show the calculation of $Q_{{\rm M}+1}$.}
    \label{FS04}
\end{figure}

Instead of \eqref{mass_a}-\eqref{mass_d}, we aim to compute
\begin{align}
    &Q_{{\rm W}+1}:=1,
    \label{mass_a'}\\
    &Q_{{\rm W}+j}:=h(\mu_{+(j-1),{\rm W}})Q_{{\rm W}+(j-1)}\quad (j=2,\cdots,\infty),
    \label{mass_b'}\\
    &Q_{{\rm W}-1}:=\sum_{j=1}^{\infty}(1-h(\mu_{+j,{\rm W}}))Q_{{\rm W}+j},
    \label{mass_c'}\\
    &Q_{{\rm W}-j}:=(1-h(\mu_{-(j-1),{\rm W}}))Q_{{\rm W}-(j-1)}\quad (j=2,\cdots,\infty),
    \label{mass_d'}
\end{align}
(see Fig.~\ref{FS04} for the illustration of this computation). Via these equations, we obtain $q_{{\rm W}j}$ by rescaling $Q_{{\rm W}j}$ as
\begin{align}
    &q_{{\rm W}j}=\frac{Q_{{\rm W}j}}{\sum_{k=\pm 1}^{\pm\infty}Q_{{\rm W}k}},
    \label{q_W}
\end{align}
which satisfies \eqref{mass_a}-\eqref{mass_d}. We should also obtain average goodnesses
\begin{align}
    &\bar{p}_{{\rm W}A}=\frac{\sum_{j=\pm 1}^{\pm\infty}Q_{{\rm W}j}\mu_{j,A}}{\sum_{j=\pm 1}^{\pm\infty}Q_{{\rm W}j}},
    \label{p_WA_ave}
\end{align}
in order to obtain Fig.~\ref{FS02}. 

In a practical computer simulation, we approximate \eqref{mass_a'}-\eqref{mass_d'} by
\begin{align}
    &\hat{Q}_{{\rm W}+1}:=1,
    \label{mass_a''}\\
    &\hat{Q}_{{\rm W}+j}:=h(\mu_{+(j-1),{\rm W}})\hat{Q}_{{\rm W}+(j-1)}\quad (j=2,\cdots,j_{\max}),
    \label{mass_b''}\\
    &\hat{Q}_{{\rm W}+j}:=0\quad (j=j_{\max}+1,\cdots,\infty),
    \label{mass_b2''}\\
    &\hat{Q}_{{\rm W}-1}:=\sum_{j=1}^{\infty}(1-h(\mu_{+j,{\rm W}}))\hat{Q}_{{\rm W}+j}=\sum_{j=1}^{j_{\max}}(1-h(\mu_{+j,{\rm W}}))\hat{Q}_{{\rm W}+j},
    \label{mass_c''}\\
    &\hat{Q}_{{\rm W}-j}:=(1-h(\mu_{-(j-1),{\rm W}}))\hat{Q}_{{\rm W}-(j-1)}\quad (j=2,\cdots,j_{\max}),
    \label{mass_d''}\\
    &\hat{Q}_{{\rm W}-j}:=0\quad (j=j_{\max}+1,\cdots,\infty),
    \label{mass_d2''}
\end{align}
with sufficient large $j_{\max}(=10^4)$ (see Fig.~\ref{FS04} for the illustration of this computation). We will show below that these computationally obtained $\hat{Q}_{{\rm W}j}$ well approximate $Q_{{\rm W}j}$. Note that in the following calculations we use the fact that 
\begin{align}
    e_2\le \mu_{j,A}\le 1-e_2
\end{align}
holds for all $j=\pm 1,\cdots, \pm\infty$ and $A$.
% FS04

From the definition, we obtain
\begin{align}
    Q_{{\rm W}+j}-\hat{Q}_{{\rm W}+j}=0\quad (j=1,\cdots,j_{\max}).
\end{align}
Then, we obtain
\begin{align}
    &Q_{{\rm W}+j}=Q_{{\rm W}+1}\prod_{k=1}^{j-1}\mu_{+k,{\rm W}}\le (1-e_2)^{j-1},\\
    &\Rightarrow \sum_{j=j_{\max}+1}^{\infty} (Q_{{\rm W}+j}-\hat{Q}_{{\rm W}+j})=\sum_{j=j_{\max}+1}^{\infty} Q_{{\rm W}+j}\le \sum_{j=j_{\max}+1}^{\infty} (1-e_2)^{j-1}=\frac{1}{e_2}(1-e_2)^{j_{\max}}.
\end{align}
Then, we have
\begin{align}
    &Q_{{\rm W}-1}=\sum_{j=1}^{\infty}Q_{{\rm W}+j}(1-\mu_{+j,{\rm W}})=\underbrace{\sum_{j=1}^{j_{\max}}Q_{{\rm W}+j}(1-\mu_{+j,{\rm W}})}_{=\hat{Q}_{{\rm W}-1}}+\sum_{j=j_{\max}+1}^{\infty}Q_{{\rm W}+j}(1-\mu_{+j,{\rm W}})\le\hat{Q}_{{\rm W}-1}+\frac{1}{e_2}(1-e_2)^{j_{\max}},\\
    &\Rightarrow Q_{{\rm W}-j}=Q_{{\rm W}-1}\prod_{k=1}^{j-1}(1-\mu_{-k,{\rm W}})=\underbrace{\hat{Q}_{{\rm W}-1}\prod_{k=1}^{j-1}(1-\mu_{-k,{\rm W}})}_{=\hat{Q}_{{\rm W}-j}}+(Q_{{\rm W}-1}-\hat{Q}_{{\rm W}-1})\prod_{k=1}^{j-1}(1-\mu_{-k,{\rm W}})\le \hat{Q}_{{\rm W}-j}+\frac{1}{e_2}(1-e_2)^{j_{\max}+j-1},\\
    &\Rightarrow\sum_{j=1}^{j_{\max}}(Q_{{\rm W}-j}-\hat{Q}_{{\rm W}-j})\le\frac{1}{e_2^2}(1-e_2)^{j_{\max}},
\end{align}
We also obtain
\begin{align}
    &Q_{{\rm W}-1}=\sum_{j=1}^{\infty}Q_{{\rm W}+j}(1-\mu_{+j,{\rm W}})\le\sum_{j=1}^{\infty}Q_{{\rm W}+j}\le \frac{1}{e_2},\\
    &\Rightarrow Q_{{\rm W}-j}=Q_{{\rm W}-1}\prod_{k=1}^{j-1}(1-\mu_{-k,{\rm W}})\le \frac{1}{e_2}(1-e_2)^{j-1},\\
    &\Rightarrow \sum_{j=j_{\max}+1}^{\infty}(Q_{{\rm W}-j}-\hat{Q}_{{\rm W}-j})=\sum_{j=j_{\max}+1}^{\infty}Q_{{\rm W}-j}\le \sum_{j=j_{\max}+1}^{\infty}\frac{1}{e_2}(1-e_2)^{j-1}=\frac{1}{e_2^2}(1-e_2)^{j_{\max}}.
\end{align}

From the above error estimations, we can obtain upper and lower bounds of \eqref{q_W} and \eqref{p_WA_ave} as
\begin{align}
    &\frac{\hat{Q}_{{\rm W}j}}{\sum_{j=\pm 1}^{\pm j_{\max}}\hat{Q}_{{\rm W}j}+\frac{3}{e_2^2}(1-e_2)^{j_{\max}}}\le q_{{\rm W}j}\le \frac{\hat{Q}_{{\rm W}j}+\frac{1}{e_2}(1-e_2)^{j_{\max}}}{\sum_{j=\pm 1}^{\pm j_{\max}}\hat{Q}_{{\rm W}j}}.\\
    &\frac{\sum_{j=\pm 1}^{\pm j_{\max}}\hat{Q}_{{\rm W}j}\mu_{j,A}}{\sum_{j=\pm 1}^{\pm j_{\max}}\hat{Q}_{{\rm W}j}+\frac{3}{e_2^2}(1-e_2)^{j_{\max}}}\le \bar{p}_{{\rm W}A}\le \frac{\sum_{j=\pm 1}^{\pm j_{\max}}\hat{Q}_{{\rm W}j}\mu_{j,A}+\frac{3}{e_2^2}(1-e_2)^{j_{\max}}}{\sum_{j=\pm 1}^{\pm j_{\max}}\hat{Q}_{{\rm W}j}}.
\end{align}
Here, we used
\begin{align}
    \hat{Q}_{{\rm W}j}\le Q_{{\rm W}j}&= \hat{Q}_{{\rm W}j}+(Q_{{\rm W}j}-\hat{Q}_{{\rm W}j})\\
    &\le \hat{Q}_{{\rm W}j}+\underbrace{\max_{j}(Q_{{\rm W}j}-\hat{Q}_{{\rm W}j})}_{=Q_{{\rm W}-1}-\hat{Q}_{{\rm W}-1}\le\frac{1}{e_2}(1-e_2)^{j_{\max}}}\le\hat{Q}_{{\rm W}j}+\frac{1}{e_2}(1-e_2)^{j_{\max}},\\
    \sum_{j=\pm 1}^{\pm j_{\max}}\hat{Q}_{{\rm W}j}\le \sum_{j=\pm 1}^{\pm\infty}Q_{{\rm W}j}&=\sum_{j=\pm 1}^{\pm j_{\max}}\hat{Q}_{{\rm W}j}+\sum_{j=\pm 1}^{\pm\infty}(Q_{{\rm W}j}-\hat{Q}_{{\rm W}j})\\
    &=\sum_{j=\pm 1}^{\pm j_{\max}}\hat{Q}_{{\rm W}j}+\underbrace{\sum_{j=1}^{j_{\max}+1}(Q_{{\rm W}j}-\hat{Q}_{{\rm W}-j})}_{\le \frac{1}{e_2^2}(1-e_2)^{j_{\max}}}+\underbrace{\sum_{j=j_{\max}+1}^{\infty}(Q_{{\rm W}+j}-\hat{Q}_{{\rm W}+j})}_{\le \frac{1}{e_2}(1-e_2)^{j_{\max}}}+\underbrace{\sum_{j=j_{\max}+1}^{\infty}(Q_{{\rm W}-j}-\hat{Q}_{{\rm W}-j})}_{\le \frac{1}{e_2^2}(1-e_2)^{j_{\max}}}\\
    &\le\sum_{j=\pm 1}^{\pm j_{\max}}\hat{Q}_{{\rm W}j}+\frac{3}{e_2^2}(1-e_2)^{j_{\max}},\\
    \sum_{j=\pm 1}^{\pm j_{\max}}\hat{Q}_{{\rm W}j}\mu_{j,A}\le \sum_{j=\pm 1}^{\pm\infty}Q_{{\rm W}j}\mu_{j,A}&\le \sum_{j=\pm 1}^{\pm j_{\max}}\hat{Q}_{{\rm W}j}\mu_{j,A}+\sum_{j=\pm 1}^{\pm\infty}(Q_{{\rm W}j}-\hat{Q}_{{\rm W}j})\\
    &\le \sum_{j=\pm 1}^{\pm j_{\max}}\hat{Q}_{{\rm W}j}\mu_{j,A}+\frac{3}{e_2^2}(1-e_2)^{j_{\max}}.
\end{align}

In the same way, let us consider \eqref{mass_e}-\eqref{mass_h} and compute
\begin{align}
    &Q_{{\rm M}+1}:=\sum_{j=1}^{\infty}h(\mu_{-j,{\rm M}})Q_{{\rm W}-j},
    \label{mass_e'}\\
    &Q_{{\rm M}+j}:=h(\mu_{+(j-1),{\rm M}})Q_{{\rm W}+(j-1)}\quad (j=2,\cdots,\infty),
    \label{mass_f'}\\
    &Q_{{\rm M}-1}:=\sum_{j=1}^{\infty}(1-h(\mu_{+j,{\rm M}}))Q_{{\rm W}+j},
    \label{mass_g'}\\
    &Q_{{\rm M}-j}:=(1-h(\mu_{-(j-1),{\rm M}}))Q_{{\rm W}-(j-1)}\quad (j=2,\cdots,\infty).
    \label{mass_h'}
\end{align}
Via these equations, we obtain $q_{{\rm W}j}$ by rescaling $Q_{{\rm M}j}$ as
\begin{align}
    q_{{\rm M}j}=\frac{Q_{{\rm M}j}}{\sum_{k=\pm 1}^{\pm\infty}Q_{{\rm M}k}},
    \label{q_M}
\end{align}
which satisfies \eqref{mass_e}-\eqref{mass_h}. We also need to obtain average goodnesses
\begin{align}
    &\bar{p}_{{\rm M}A}=\frac{\sum_{j=\pm 1}^{\pm\infty}Q_{{\rm M}j}\mu_{j,A}}{\sum_{j=\pm 1}^{\pm\infty}Q_{{\rm M}j}},
    \label{p_MA_ave}
\end{align}
in order to obtain Fig.~\ref{FS02}. 

In a practical computer simulation, we approximate \eqref{mass_e'}-\eqref{mass_h'} by
\begin{align}
    &\hat{Q}_{{\rm M}+1}:=\sum_{j=1}^{\infty}h(\mu_{-j,{\rm M}})\hat{Q}_{{\rm W}-j}=\sum_{j=1}^{j_{\max}}h(\mu_{-j,{\rm M}})\hat{Q}_{{\rm W}-j},
    \label{mass_a'''}\\
    &\hat{Q}_{{\rm M}+j}:=h(\mu_{+(j-1),{\rm M}})\hat{Q}_{{\rm W}+(j-1)}\quad (j=2,\cdots,j_{\max}),
    \label{mass_b'''}\\
    &\hat{Q}_{{\rm M}+j}:=0\quad (j=j_{\max}+1,\cdots,\infty),
    \label{mass_b2'''}\\
    &\hat{Q}_{{\rm M}-1}:=\sum_{j=1}^{\infty}(1-h(\mu_{+j,{\rm M}}))\hat{Q}_{{\rm W}+j}=\sum_{j=1}^{j_{\max}}(1-h(\mu_{+j,{\rm M}}))\hat{Q}_{{\rm W}+j},
    \label{mass_c'''}\\
    &\hat{Q}_{{\rm M}-j}:=(1-h(\mu_{-(j-1),{\rm M}}))\hat{Q}_{{\rm W}-(j-1)}\quad (j=2,\cdots,j_{\max}),
    \label{mass_d'''}\\
    &\hat{Q}_{{\rm M}-j}:=0\quad (j=j_{\max}+1,\cdots,\infty),
    \label{mass_d2'''}
\end{align}
with sufficient large $j_{\max}(=10^4)$.

By exactly similar calculations, we obtain
\begin{align}
    &Q_{{\rm M}+j}-\hat{Q}_{{\rm M}+j}=0\quad (j=2,\cdots,j_{\max}),\\
    &\sum_{j=j_{\max+1}}^{\infty}(Q_{{\rm M}+j}-\hat{Q}_{{\rm M}+j})\le \frac{1}{e_2}(1-e_2)^{j_{\max}},\\
    &\sum_{j=1}^{j_{\max}}(Q_{{\rm M}-j}-\hat{Q}_{{\rm M}-j})\le \frac{1}{e_2^2}(1-e_2)^{j_{\max}},\\
    &\sum_{j=j_{\max+1}}^{\infty}(Q_{{\rm M}-j}-\hat{Q}_{{\rm M}-j})\le \frac{1}{e_2^2}(1-e_2)^{j_{\max}}.
\end{align}
The difference from $Q_{{\rm W}j}$ exists only in $j=+1$, as
\begin{align}
    Q_{{\rm M}+1}&=\sum_{j=1}^{\infty}Q_{{\rm W}-j}\mu_{-j,{\rm M}}=\underbrace{\sum_{j=1}^{j_{\max}}\hat{Q}_{{\rm W}-j}\mu_{-j,{\rm M}}}_{=\hat{q}_{{\rm M}+1}}+\underbrace{\sum_{j=1}^{j_{\max}}(Q_{{\rm W}-j}-\hat{Q}_{{\rm W}-j})\mu_{-j,{\rm M}}}_{\le \frac{1}{e_2^2}(1-e_2)^{j_{\max}}}+\underbrace{\sum_{j=j_{\max}+1}^{\infty}\hat{Q}_{{\rm W}-j}\mu_{-j,{\rm M}}}_{\le \frac{1}{e_2^2}(1-e_2)^{j_{\max}}}\\
    &\le \hat{Q}_{{\rm M}+1}+\frac{2}{e_2^2}(1-e_2)^{j_{\max}},
\end{align}
(see the area surrounded by dots in Fig.~\ref{FS04} for the illustration of this computation).

From the above error estimations, we can obtain upper and lower bounds of \eqref{q_M} and \eqref{p_MA_ave} as
\begin{align}
    &\frac{\hat{Q}_{{\rm M}j}}{\sum_{j=\pm 1}^{\pm j_{\max}}\hat{Q}_{{\rm M}j}+\frac{5}{e_2^2}(1-e_2)^{j_{\max}}}\le q_{{\rm M}j}\le \frac{\hat{Q}_{{\rm M}j}+\frac{2}{e_2^2}(1-e_2)^{j_{\max}}}{\sum_{j=\pm 1}^{\pm j_{\max}}\hat{Q}_{{\rm M}j}},\\
    &\frac{\sum_{j=\pm 1}^{\pm j_{\max}}\hat{Q}_{{\rm M}j}\mu_{j,A}}{\sum_{j=\pm 1}^{\pm j_{\max}}\hat{Q}_{{\rm M}j}+\frac{5}{e_2^2}(1-e_2)^{j_{\max}}}\le \bar{p}_{{\rm M}A}\le \frac{\sum_{j=\pm 1}^{\pm j_{\max}}\hat{Q}_{{\rm M}j}\mu_{j,A}+\frac{5}{e_2^2}(1-e_2)^{j_{\max}}}{\sum_{j=\pm 1}^{\pm j_{\max}}\hat{Q}_{{\rm M}j}}.
\end{align}
Here, we used
\begin{align}
    \hat{Q}_{{\rm M}j}\le Q_{{\rm M}j}&=\hat{Q}_{{\rm M}j}+(Q_{{\rm M}j}-\hat{Q}_{{\rm M}j})\\
    &\le \hat{Q}_{{\rm M}j}+\underbrace{\max_{j}(Q_{{\rm M}j}-\hat{Q}_{{\rm M}j})}_{=Q_{{\rm M}+1}-\hat{Q}_{{\rm M}+1}\le\frac{1}{e_2^2}(1-e_2)^{j_{\max}}}\le \hat{Q}_{{\rm M}j}+\frac{1}{e_2^2}(1-e_2)^{j_{\max}},\\
    \sum_{j=\pm 1}^{\pm j_{\max}}\hat{Q}_{{\rm M}j}\le\sum_{j=\pm 1}^{\pm\infty} Q_{{\rm M}j}&=\sum_{j=\pm 1}^{\pm j_{\max}}\hat{Q}_{{\rm M}j}+\underbrace{(Q_{{\rm M}+1}-\hat{Q}_{{\rm M}+1})}_{\le \frac{2}{e_2^2}(1-e_2)^{j_{\max}}}+\underbrace{\sum_{j=1}^{j_{\max}}(Q_{{\rm M}-j}-\hat{Q}_{{\rm M}-j})}_{\le \frac{1}{e_2^2}(1-e_2)^{j_{\max}}}\\
    &\quad+\underbrace{\sum_{j=j_{\max}+1}^{\infty}(Q_{{\rm M}+j}-\hat{Q}_{{\rm M}+j})}_{\le \frac{1}{e_2}(1-e_2)^{j_{\max}}}+\underbrace{\sum_{j=j_{\max}+1}^{\infty}(Q_{{\rm M}-j}-\hat{Q}_{{\rm M}-j})}_{\le \frac{1}{e_2^2}(1-e_2)^{j_{\max}}}\\
    &\le\sum_{j=\pm 1}^{\pm j_{\max}}\hat{Q}_{{\rm M}j}+\frac{5}{e_2^2}(1-e_2)^{j_{\max}},\\
    \sum_{j=\pm 1}^{\pm j_{\max}}\hat{Q}_{{\rm M}j}\mu_{j,A}\le\sum_{j=\pm 1}^{\pm\infty} Q_{{\rm M}j}\mu_{j,A}&\le\sum_{j=\pm 1}^{\pm j_{\max}}\hat{Q}_{{\rm M}j}\mu_{j,A}+\sum_{j=\pm 1}^{\pm\infty}(Q_{{\rm M}j}-\hat{Q}_{{\rm M}j})\\
    &\le \sum_{j=\pm 1}^{\pm j_{\max}}\hat{Q}_{{\rm M}j}\mu_{j,A}+\frac{5}{e_2^2}(1-e_2)^{j_{\max}}.
\end{align}

\end{document}